\documentclass[a4paper, fleqn, usenatbib, useAMS]{mnras}

\def\mcut{M_{\rm cut}}
\def\zh{\zeta_{\rm h}}
\def\tm{\theta_{\rm m}}
\def\DL{\Delta\mathscr{L}}
\def\textt{\texttt}

\def\aap{AA}
\def\apjl{ApJL}
\def\apjs{ApJS}
\def\mnras{MNRAS}

\usepackage{enumitem}
\usepackage{mathrsfs}
\usepackage{afterpage}
\usepackage{multicol}        
\usepackage{graphicx}
\usepackage{float}
\usepackage{bm} 
\usepackage{amssymb}
\usepackage{hyperref}
\usepackage{amsmath} 
\usepackage{subfig} 
\usepackage{pdflscape}
\usepackage{natbib}
\usepackage[export]{adjustbox}
\usepackage{mathrsfs} 
\usepackage{mathtools}

\title[Halo detectability in warm dark matter models]{Halo concentration strengthens dark matter constraints in galaxy-galaxy strong lensing analyses}
\author[N. C. Amorisco et al.]{Nicola C. Amorisco$^{1}$\thanks{E-mail:
nicola.c.amorisco@durham.ac.uk}\thanks{Ernest Rutherford Fellow},  James Nightingale$^{1}$, Qiuhan He$^{1}$, \newauthor
Aristeidis Amvrosiadis$^{1}$, 
 Xiaoyue Cao$^{3,2}$,
Shaun Cole$^{1}$, Amy Etherington$^{1}$, \newauthor
Carlos S. Frenk$^{1}$, Ran Li$^{2,3}$, Richard Massey$^{1}$, 
 Andrew Robertson$^{1}$ 
\\
$^{1}$Institute for Computational Cosmology, Department of Physics, Durham University, South Road, Durham DH1 3LE, UK\\
$^{2}$National Astronomical Observatories, Chinese Academy of Sciences, 20A Datun Road,Chaoyang District, Beijing 100012, China \\
$^{3}$School of Astronomy and Space Science, University of Chinese Academy of Sciences, Beijing 100049, China\\}

\begin{document}


\pagerange{\pageref{firstpage}--\pageref{lastpage}} \pubyear{2021}

\maketitle

\label{firstpage}

\begin{abstract}

 \noindent A defining prediction of the cold dark matter (CDM) cosmological
  model is the existence of a very large population of low-mass
  haloes. This population is absent in models in which the dark matter
  particle is warm (WDM).  These alternatives can, in principle, be
  distinguished observationally because halos along the line-of-sight
  can perturb galaxy-galaxy strong gravitational lenses. Furthermore,
  the WDM particle mass could be deduced because the cut-off in their
  halo mass function depends on the mass of the particle. We
  systematically explore the detectability of low-mass haloes in WDM
  models by simulating and fitting mock lensed images.  Contrary to
  previous studies, we find that halos are harder to detect when they
  are either behind {\it or in front of} the
  lens. Furthermore, we find that the perturbing effect of haloes
  increases with their concentration: detectable haloes are
  systematically high-concentration haloes, and accounting for the
  scatter in the mass-concentration relation boosts the expected
  number of detections by as much as an order of magnitude. Haloes
  have lower concentration for lower particle masses and this further
  suppresses the number of detectable haloes beyond the reduction
  arising from the lower halo abundances alone. Taking these effects
  into account can make lensing constraints on the value of the mass
  function cut-off at least an order of magnitude more stringent than
  previously appreciated.

\end{abstract}


\section{Introduction}
The nature and identity of dark matter (DM) remain fundamental
open questions in contemporary astrophysics; enormous effort is
currently being directed at finding the answer. 
Numerical simulations of the cosmological process of structure
formation \citep[e.g.][]{Davies1985,Springel2005,Frenk2012} have shown
that a model based on the assumption that the DM consists of 
cold dark matter (CDM) particles can very successfully reproduce a number of large-scale astrophysical measurements 
\citep[e.g.][]{Planck2018, Wang2016, Alam2017}. 
Several plausible DM candidates behave like CDM on large scales, 
but luckily, their different physical properties can make them
distinguishable on subgalactic scales. 
The defining property of standard CDM is the nearly scale-invariant primordial power-spectrum 
of density fluctuations, which results in an equally distinctive halo
mass function, characterized by a large 
population of haloes down to masses comparable to the Earth's mass \citep{Jenkins2001, Diemand2008, 
Angulo2012, Green2005, Wang2020}. Most alternative DM models predict a
suppression of the primordial power spectrum on small scales and an
associated truncation of the halo mass function at a mass, $\mcut$. For example, in the
popular warm dark matter (WDM) model, free-streaming arising from the
thermal velocities of the particles at
early times is the cause of the suppression which occurs at a mass
scale that is roughly inversely proportional to the WDM particle mass 
\citep[e.g.][]{Avila2003, Schneider2012, Lovell2012, Bose16}.

Current constraints on the warm DM model stem primarily from a
combination of the abundance of satellite galaxies in the Milky Way
\citep{Kennedy2014,Lovell2012,Lovell2016,Newton2020} and the properties of the 
Lyman-$\alpha$ forest inferred from high-redshift QSO spectra 
\citep{Viel2013, Baur2016,Irsic2017}.  A joint analysis of these,
together with 
constraints from gravitational lensing (see below), 
place $\mcut$ at  
$\approx 4.3 \times 10^{7}$M$_\odot$ for a thermal WDM relic. These bounds, however, are
subject to possible systematics such as uncertainties in the galaxy
formation physics in the case of satellites \citep{Newton2020} and 
assumptions on the thermal history of the intergalactic medium at
high-redshift in the case of the Lyman-$\alpha$ forest
\citep[e.g.][]{Garzilli2017,Garzilli2019}.  Constraints from
independent probes, such as we discuss here, are therefore a priority.

Strong gravitational lensing has emerged as an independent way to quantify the abundance 
of low-mass DM haloes and thus constrain the WDM particle mass. 
This technique uses galaxy-galaxy strong gravitational lenses \citep[e.g.,][]{Bolton2005, Shu2016} 
to detect low-mass haloes through the perturbations they cause to the
lens image 
\citep[see also the alternative approach based on flux ratio anomalies of lensed quasars, e.g.,][]{Xu2015, Gilman2017, Gilman2019, Gilman2020, Harvey2020}.  
These perturbations make it possible to detect both satellite haloes
in the main lens (subhaloes) and low-mass `central' haloes 
along the line-of-sight (LOS) \citep{Li2016, Despali2018}, even if
they contain negligible baryonic mass. In fact, in the mass range of
interest, $\lesssim 10^8$M$_\odot$, haloes are too small to have made
a galaxy and so are completely dark \citep{Benitez-Llambay2020}. This is a great advantage as the 
abundance and structure of isolated DM haloes is unaffected by
complications associated with baryonic processes and
is very robustly determined by cosmological simulations. Distortions
of strong lenses are therefore 
an especially clean way to probe the DM particle mass. 
 
A small number of detections have already been claimed, albeit for
subhaloes more massive than those that can test WDM models \citep{Vegetti2010, Vegetti2012, Hezaveh2016}. The most challenging aspect of lensing lies in using these detections -- and 
non-detections as well -- to extract quantitative inferences about the DM model \citep[e.g.][]{Vegetti2014, Vegetti2018, Ritondale2019}. 
To do so, it is necessary to formulate robust predictions. For
example, how many detections are to be expected in a
given lensing system assuming CDM, or as a function of the WDM particle mass? 

Quantifying the number of detectable haloes, $N_{\rm d}$, in a
specific lens means identifying which DM haloes,  
out of the cosmological population of haloes, can cause `observable' perturbations to that system.
More specifically, for a warm DM particle with cutoff mass, $M_{\rm cut}$,
\begin{equation}
N_{\rm d}(M_{\rm cut}) =  \int n(x,y,z,\zh|M_{\rm cut})\cdot p(x,y,z,\zh |\theta, \textbf{n}) \ dV \bf d\zh,
\label{ndetectable}
\end{equation}
where $n$ is the cosmological number density of DM
haloes\footnote{Here we focus on LOS haloes. The same formalism
  applies to subhaloes in the main lens, albeit with a different
  density, $n$.} at sky-projected location, $(x,y)$,
redshift, $z$, and with properties, $\zh$ (such as mass, concentration, etc.), while $p$ is the probability 
of actually detecting such haloes were they to be truly present in the observed system, given the properties of the lens
itself, $\theta$, and those of the data -- for instance, the noise
properties, $\bf n$. In other words, were a halo to 
be truly present: 
\begin{itemize}
\item{$p=0$ if its perturbations are too small to be observable,
    implying that a perturbing halo mass component would not be required in the modelling to describe the data; }
\item{$p=1$ if its perturbations make a model including a perturbing mass component 
statistically preferable to one that does not.}
\end{itemize}
The increase in Bayesian evidence between the two models (or the increase in log-likelihood) is often 
used as a deciding metric, and most studies \citep[e.g.][]{Vegetti2014,Vegetti2018} have indeed reduced 
$p$ to a binary classification: were the halo to be truly present,
$p=1$, if including a perturbing mass component 
causes the Bayesian evidence or log-likelihood to increase beyond some given threshold.
This is usually referred to as the {\it sensitivity function} and, simply put, it identifies the region of 
parameter space (comprising both physical cosmological volume and halo properties) which can be actually 
probed by lensing. In contrast to the cosmological number
density of DM haloes, $n$, the sensitivity function itself 
does not directly depend on the DM model\footnote{Although it does
  depend on the density profile of the perturbing haloes, which, in
  turn, may itself depend on the DM model.}, but it does shape
expectations for the number of detectable haloes, 
$N_{\rm d}$ -- as well as expectations for any other observable
obtained through structure lensing studies.

While advanced tools to model optical strong lensing data have been
developed \citep[e.g.][]{Vegetti2009, Nightingale2021}, it remains
computationally expensive to calculate the sensitivity function and
formulate these predictions. 
Systematic exploration is required to establish the range of
properties that make a perturber detectable. 
A minimum list of the independent variables include the halo mass and halo concentration,
the projected location of the halo with respect to the lensing system, and its redshift. In addition, 
the sensitivity function is unique to each individual lensing system because degeneracies in the lensing 
effects are such that different lensing configurations can `reabsorb'
the perturbations of identical DM haloes  
with different efficiencies. In practice, mapping the entire parameter space for each lens is  
often computationally prohibitive, and a number of simplifying assumptions have been used to
obtain estimates of the integral in  equation~(\ref{ndetectable}).
Here, we explore the effect of these simplifications.
Among the independent variables mentioned above, halo concentration
and halo redshift are the most important. 

For instance, \citet{Minor2021} has recently shown that the halo
concentration must be included as a free parameter when modelling a
perturber: if the concentration is fixed, the inferred perturber's mass may be 
biased by a factor of up to 6. They also show that higher halo concentrations make perturbers more easily detectable, as the lensing effect of any mass 
distribution is driven by its surface density. However, the intrinsic scatter in the concentration of 
DM haloes \citep[e.g.][]{Neto2007, Ludlow2016, Wang2020} has so far
been ignored in sensitivity mapping studies; instead, the mass-concentration relation 
has been collapsed onto the concentration axis entirely, forcing all haloes onto the mean value for their mass. 
Additionally, the dependence of the mean mass-concentration relation on the DM model itself has also
been neglected \citep[e.g.][]{Vegetti2018,Ritondale2019}. This latter
assumption leaves cosmological halo abundances as the single measure to differentiate 
between WDM models of different particle mass, despite the fact that
warmer DM models produce 
haloes that are increasingly less dense than their equal-mass CDM
counterparts \citep[e.g.][]{Lovell2012, Bose16}.

As regards the perturber's redshift, this axis has often been
collapsed by adopting a one-to-one scaling relation 
that recasts a halo's redshift in terms of its effective mass
\citep{Li2016,Despali2018}. This is obtained by requiring that the lensing convergence
-- i.e.\ the strength of the lensing effect -- should remain nearly
constant. This is not equivalent to performing a full non-linear
search, as done on real data, and therefore does not fully take into
account the modelling degeneracies that can occur  in the real case. Lastly, we also briefly reflect on the influence of 
noise properties and the role of a specific noise realization.

To explore the parameter space of a sensitivity function fully, we use mock data and models that are somewhat simpler 
than those employed in state-of-the-art lens modelling studies \citep[e.g.][]{Vegetti2009, Nightingale2019, Powell2020}. 
Specifically, we use models featuring parametric sources rather than non-parametric pixelized sources 
\citep[e.g.][]{Warren2003, Dye2005,  Birrer2015a, Nightingale2015}. This allows us to develop and apply a 
fitting procedure based on a gradient descent algorithm, which is efficient enough to enable the exploration
of all of the independent dimensions of the parameter space relevant to this problem. This work compares how 
previous assumptions regarding the sensitivity function affect the
power of strong lensing to discriminate between different 
DM models, which we assume is not strongly dependent on the specific approach to source modelling.

In this work, we concentrate on LOS isolated perturbers, which we
model as pure NFW profiles \citep{Navarro1997}. Satellite subhaloes have different density profiles and their number
density in the main lens is affected by a variety of physical
processes. He et al. (in preparation) use the high-resolution
cosmological hydrodynamical simulations of \cite{Richings2021} to
facilitate a comparison between the lensing perturbations caused by
satellite and LOS halos, thereby allowing an estimate of their
relative importance. In their study, they also employ different sets of lens configurations, modelling and fitting techniques. While they do not focus on the effect of halo concentration, their independent analysis finds quantitatively similar results on the redshift dependence of a perturber's detectability, which further reinforces the need to move away from the approximations used so far. 

We stress that the present work is not intended as a substitute for
analyses aimed at quantifying the sensitivity function of actual sets
of observed lenses -- which should be tailored to the lens
configurations featured in the real data and should be performed using
the same modelling techniques as applied to the real data.

This paper is organised as follows.
Section~2 provides a quick overview of the standard procedure used to estimate the sensitivity function;
Section~3 describes our modelling framework and fitting procedures;
Section~4 describes our results, focusing on the dependence of halo detectability on redshift and concentration;
Section~5 uses our sensitivity maps to estimate the number of expected
detections for different warm DM models; and 
Section~6 examines the consequences of our results for future substructure lensing studies.

\section{Sensitivity Mapping Overview}
In the interest of clarity, we start by outlining the procedure usually adopted to measure the sensitivity function. 
Let us assume, for example, that we wish to predict the number of
detectable haloes~(equation~\ref{ndetectable}) for a specific strong
lens. One would start by  modelling the lens image in order to infer: (i) a mass model for the lens galaxy; (ii) a light model for the source galaxy. Then,

\begin{enumerate}[label=\arabic*)]
\item{using these and the noise properties of the data themselves, one
    {simulates a strong lens image which includes}  
a perturbing DM halo with a set of properties (such as mass
and concentration), located at projected location, $(x,y)$
and redshift, $z$;}
\item{one  fits these mock data in two full but distinct non-linear
    searches, the first with a model that includes a perturbing halo mass component, the second  without.} 
\item{one compares the two model fits, by means of the
    Bayesian evidence or the maximum log-likelihood. If a model
    including a halo mass component provides a significantly better
    fit, the original data are sensitive to a halo with those specific
    properties, thereby mapping the probability of detection, $p$.}
\end{enumerate}

\noindent This procedure is repeated multiple times so as to sample the entire parameter space of perturbers' locations and properties. 

We dwelve more deeply into the different steps of this procedure in
the next section. In particular, we shall demonstrate that, in
practice, we do not need to perform one of the fits at all.

\section{Modelling framework}

We assume we have optical (mock) data, $\textbf{d}$, for a lensing system characterized by the presence of some perturbing 
LOS halo with properties, $\bf\zh$, and we wish to assess its detectability\footnote{For compactness of notation, we include the
perturber's sky coordinates, $(x,y)$, and redshift, $z$, in the halo
properties, $\zh$.}. We do so by quantifying the
log-likelihood\footnote{By `log-likelihood' we mean the natural
  logarithm of the likelihood. All other instances of `log' in this
  work represent the base 10 logarithm.}
difference
\begin{equation}
\DL \left( \bf\zh\right) = \mathscr{L}_{\rm m,h}\left(\bf{\hat\theta}_{\rm m}, \bf{\hat\zeta}_{\rm h}\right) - \mathscr{L}_{\rm m}\left(\bf{\bar\theta}_{\rm m}\right),
\label{DL}
\end{equation}
where:
\begin{itemize}
\item{$\mathscr{L}_{\rm m}\left(\bf{\bar\theta}_{\rm m}\right)$ is the log-likelihood value corresponding to the best-fitting model that does not include a 
perturbing halo mass component. This model is optimized over the parameters of the so called macromodel alone, $\bf{\theta}_{\rm m}$,
which include the parameters describing the source, $\bf{\theta}_{\rm
  s}$, and those describing the lens, as well as any external shear, $\bf{\theta}_{\rm l}$;}
\item{$\mathscr{L}_{\rm m,h}\left(\bf{\hat\theta}_{\rm m}, \bf{\hat\zeta}_{\rm h}\right)$ is the log-likelihood value corresponding to the best-fitting model 
that does include a perturbing halo mass component, which is optimized over both macromodel and halo parameters,
with best-fitting values, $\bf{\hat\theta}_{\rm m}$ and $\bf{\hat\zeta}_{\rm h}$ respectively.}
\end{itemize}
We take the detection probability, $p(\zh)$, in
equation~(\ref{ndetectable}) to be a function of the log-likelihood gain, $\DL$:
\begin{equation}
p(\zh) = p(\DL(\zh)), 
\label{pDL}
\end{equation}
that is, perturbing haloes that result in higher values of the
log-likelihood gain, $\DL$, are more easily detectable; 
for the moment we defer fixing the functional link between detection probability and log-likelihood increase.

All likelihood values are obtained by comparing the pixelised flux values, $\bf d$,
with the model flux distributions, $\bf f$. We ignore the effect of noise covariance (due to effects like PSF convolution) so that we simply have
\begin{equation}
\mathscr{L} = - {1\over 2} \sum_{\rm pixels} \Big|  {{{\bf d - \bf f}\over \bf n}} \Big|^2 \ ,
\label{calL}
\end{equation}
where $\bf n$ represents the noise map associated with the data and we
can neglect the normalization term in $\ln\bf n$ because 
we are only interested in log-likelihood differences. We assume that the data only include flux from the source, 
i.e.\ that both sky background and lens fluxes have been subtracted before performing the fits. 

A number of authors \citep[e.g.][]{Vegetti2018,Ritondale2019} have
argued for adopting the gain in Bayesian evidence, rather than in the log-likelihood, 
as a basis for quantifying halo detectability. 
We recall that the evidence is defined as the integral of the posterior over the entire parameter space.
We agree that the gain in log evidence is a more sound statistical metric for model 
comparison. However, calculating the evidence is orders of magnitude more computationally expensive than identifying the best-fitting model, as it requires  sampling the likelihood surface over the entire parameter space in order to integrate it. This has  become one of the main reasons behind the need for making simplifications when calculating 
the sensitivity function.

It is important to stress that, as long as the same criterion is consistently employed to 
both (i) detect perturbers on real data and (ii) measure the sensitivity function and make predictions for the expected number of detections, an evidence based strategy and a likelihood based one are both
perfectly acceptable and will both yield correct results for the DM particle mass.  
It is indeed possible that an evidence-based criterion may help to weed out false detections better in studies of real data. On the other hand, this is not strictly necessary
{if the models used in sensitivity mapping share the same complexity of the real data, and therefore also share any spurious detections}. In fact, a computationally more efficient criterion may facilitate a robust characterization of the properties and frequency of false positives, and therefore help to take them into account when 
making predictions. 
For the present purposes, 
a likelihood-based criterion is beneficial in that it allows us to explore the entire parameter space systematically. Furthermore, for data of sufficiently high quality, the gain in evidence and in log-likelihood
become equivalent \citep[{as embodied by the BIC,} see e.g.][]{ESLII, ICSM}, and numerical experiments seem to indicate that
the quality of {\it Hubble Space Telescope} (HST) data is, in fact, high enough for the two approaches to often provide very 
similar results (He et al.\ in preparation).

\subsection{Model families and mock data}

Our lensing systems feature:
\begin{itemize}
\item{a power-law mass profile to represent the main lens \citep{Tessore2016}, characterized by the following free parameters: 
$(x_{\rm l}, y_{\rm l})$, the centre of the mass distribution;
$\epsilon_{\rm l}$,  the Einstein radius; $\beta_{\rm l}$,  
the slope of the mass profile; $(e_{1, l}, e_{2, l})$, the two
independent components of the profile's ellipticity;  
$(\gamma_1,\gamma_2)$, the two independent components of the external shear.}
\item{a parametric source with a Sersic profile: projected
    centre, $(x_{\rm s}, y_{\rm s})$; effective radius, $r_{\rm eff}$; 
ellipticity, $(e_{1,s}, e_{2, s})$; Sersic index, $n_{\rm s}$; and total flux, $I_{\rm s}$.}
\end{itemize}
These two components define our macromodel: $\tm$ is therefore a 15-dimensional vector.

The perturbing haloes are modelled with spherically symmetric
Navarro-Frenk-White mass profiles \citep{Navarro1997},  
introducing the following additional 5 parameters: projected centre,
$(x_{\rm h}, y_{\rm h})$; redshift, $z_{\rm h}$; mass,  $M_{\rm h}$; 
and concentration $c_{\rm h}$. Throughout the paper we take halo
masses, $M_{\rm h}$, to be the virial mass, $M_{200}$, i.e. the mass
contained within a sphere of density 200 times the critical density.
We use the open source software \texttt{PyAutoLens}\footnote{{\tt PyAutoLens} 
is open-source and available from \url{https://github.com/Jammy2211/PyAutoLens}} \citep{Nightingale2015, Nightingale2018, Nightingale2021} to generate all of our mock data and for all 
our lensing modelling. 

\begin{figure}
\centering
\includegraphics[width=\columnwidth]{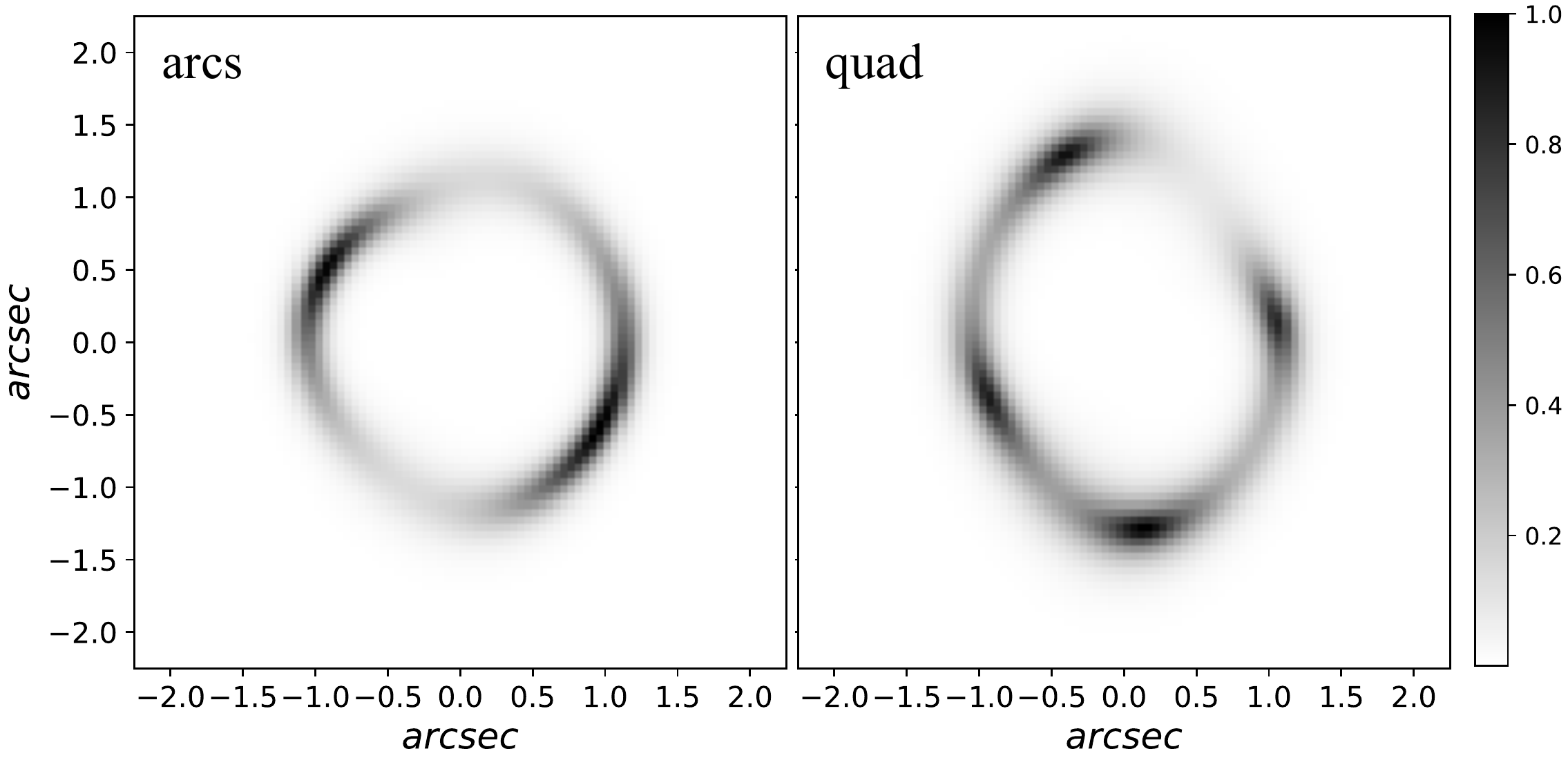}
\caption{Model fluxes for the two lensing configurations used in this
  study, giant arcs (left), quad (right) (linear scale with arbitrary units).}
\label{macromodelsfig}
\end{figure}

Within this framework, we choose two different lensing configurations for our exploration. Both have their source
at $z_{\rm s} =1$ and lens at $z_{\rm l}=0.5$, but one is in a quad-configuration while the other features two asymmetric arcs. 
Values for the ground truth macromodel parameters, $\tm$, are recorded
in Table~\ref{inputmacro}; Fig.~\ref{macromodelsfig} illustrates their
geometry, displaying the corresponding model fluxes,
$\bf{f}(\tm)$. The exact values are of no importance here: we simply choose 
sets of parameters that are qualitatively in line with what is found in lensing studies of real systems, 
and we use a pair of different configurations to ensure that the trends we identify are not a peculiarity of the specific 
system we happen to adopt. Throughout this work, we use pixel size and point-spread function (PSF) width typical for HST data, 
fixing both quantities at $0.05''$.

\subsection{Signal-to-noise and noise realization}
\label{stonoise}

The sensitivity function of a lens scales with the quality of the
available data, quantified here by the noise map, $\bf n$, 
and the associated maximum value of the signal-to-noise, $\textbf{n}\propto 1/SN_{\rm max}$.
Notice, however, that while the noise map itself is known, the actual noise realization of the observed data is not. 
We are therefore interested in assessing how different noise realizations affect 
the value of the log-likelihood change.

Let us assume that the data are characterized by a noise realization, $\bf r$, so that $\bf d = \langle d\rangle+r$, where $\langle\cdot\rangle$ denotes an average over noise realizations. In the case of mock data, $\langle \bf d\rangle$ is the input model flux corresponding to the ground truth model 
parameters, which we indicate with $(\theta_{\rm m},\zh)$: $\langle {\bf d}\rangle=\textbf{f}(\theta_{\rm m},\zh)$. The log-likelihood gain is as in equation~(\ref{DL}), with terms in the same order:
\begin{equation}
\DL \left( \bf\zh, \textbf{r}\right) = - {1\over 2} \Big|  {{{\textbf{d} - \textbf{f}(\hat\theta_{\rm m}, \hat\zh)}\over \bf n}} \Big|^2 + {1\over 2} \Big|  {{{\textbf{d} - \textbf{f}(\bar\theta_{\rm m})}\over \bf n}} \Big|^2,
\label{expand1}
\end{equation}
where we have implicitly assumed a sum over image pixels. The fluxes, $\textbf{f}(\hat\theta_{\rm m}, \hat\zh)$, and $\textbf{f}(\bar\theta_{\rm m})$, correspond to the model that best fit the noise-corrupted data, $\bf d$, with and without an extra halo.  
It is convenient to consider instead the 
model fluxes that provide the best fit to the noise-free data, 
$\langle\bf d\rangle$, which we refer to as $\textbf{f}_{\rm h}$ and $\textbf{f}_{\rm m}$. These models do not achieve the maximum log-likelihood values we require in equation~(\ref{expand1}). For example, for the model including a halo mass component, 
\begin{equation}
{1\over 2} \Big|  {{{\textbf{d} - \textbf{f}(\hat\theta_{\rm m}, \hat\zh)}\over \bf n}} \Big|^2 = {1\over 2} \Big|  {{{\textbf{d} - \textbf{f}_{\rm h}}\over \bf n}} \Big|^2-l_{\rm h}^{\rm bf}(\textbf{r})\ ,
\label{lbfshift1}
\end{equation}
where the difference, $l_{\rm h}^{\rm bf}$, is a function of the noise realization and has a positive value. 
Similarly, 
\begin{equation}
{1\over 2} \Big|  {{{\textbf{d} - \textbf{f}(\bar\theta_{\rm m})}\over \bf n}} \Big|^2 = {1\over 2} \Big|  {{{\textbf{d} - \textbf{f}_{\rm m}}\over \bf n}} \Big|^2-l_{\rm m}^{\rm bf}(\textbf{r})\ .
\label{lbfshift2}
\end{equation}
Furthermore, in order to highlight the dependence on the noise realization, $\bf r$, we can recast the model fluxes, $\textbf{f}_{\rm h}$
 and $\textbf{f}_{\rm m}$, in terms of their associated residuals, $\textbf{f}_{\rm h} = \langle \textbf{d}\rangle+\delta_{\rm h}$, and $\textbf{f}_{\rm m}= \langle \textbf{d}\rangle+\delta_{\rm m}$, 
which gives:
\begin{equation}
{1\over 2} \Big|  {{{\textbf{d} - \textbf{f}(\hat\theta_{\rm m}, \hat\zh)}\over \bf n}} \Big|^2 = {1\over 2} \Big|  {{{\textbf{r} - {\delta}_{\rm h}}\over \bf n}} \Big|^2-l_{\rm h}^{\rm bf}(\textbf{r}),
\label{lbfshift1a}
\end{equation}
and 
\begin{equation}
{1\over 2} \Big|  {{{\textbf{d} - \textbf{f}(\bar\theta_{\rm m})}\over \bf n}} \Big|^2 = {1\over 2} \Big|  {{{\textbf{r} - {\delta}_{\rm m}}\over \bf n}} \Big|^2-l_{\rm m}^{\rm bf}(\textbf{r}).
\label{lbfshift2a}
\end{equation}
Equation~(\ref{expand1}) is therefore the difference between the
right-hand sides of the two equations  above. We are interested in the mean and the standard deviation of this difference across noise realizations. 

Let us first consider the two shifts, $l^{\rm bf}_{\rm h}$, and $l_{\rm m}^{\rm bf}$. These are nonzero, but they are not the leading terms of equation~(\ref{expand1}) that we seek. 
We can show this by estimating their average magnitude across varying
noise realizations. This can be calculated analytically under the assumption
that the likelihood surface is Gaussian. If so, we can see that 
\begin{equation}
\langle l_{\rm bf}\rangle = {k\over 2}\ln 2,
\label{lbfshift3}
\end{equation}
which is valid for both $l^{\rm bf}_{\rm h}$ and $l_{\rm m}^{\rm bf}$, and in which $k$ is the number of independent parameters in the likelihood. In our case, the model which 
includes a halo mass component features 5 additional free parameters, so that 
\begin{equation}
\langle l^{\rm bf}_{\rm h}-l^{\rm bf}_{\rm m}\rangle = {5\over 2}\ln 2 \approx 1.73.
\label{lbfshift4}
\end{equation}
This is considerably smaller than the log-likelihood differences we are after and we therefore ignore these terms from now on.

By expanding the chi-square
terms in Eqns.~(\ref{lbfshift1a}) and~(\ref{lbfshift2a}), we
finally obtain the leading terms we are interested in:
\begin{equation}
\DL \left( \bf\zh, \textbf{r} \right) \approx {1\over 2} \Big(\Big|{\delta_{\rm h}\over{\bf n}} \Big|^2 - \Big|{\delta_{\rm m}\over{\bf n}} \Big|^2\Big) + {{\textbf{r}}\over{\textbf{n}}}\cdot{{\delta_{\rm h}-\delta_{\rm m}}\over{\textbf{n}}}.
\label{expandall}
\end{equation}
%
Here, the first term quantifies the inability of a model that does not include a perturbing halo mass component
to describe the perturbed data. This is what sensitivity mapping is after, and  is independent of the noise realization. 
The second term introduces scatter in the measurement of 
the log-likelihood gain as a consequence of varying noise realizations, $\bf r$. The case we are interested in 
is the one in which a model featuring a halo mass component provides a
satisfactory fit to the data,  
$\delta_{\rm h}/\textbf{n}\ll 1$. This is also the case of mock data in which 
the model used to generate data is the same used to fit it: $\langle
\bf d\rangle = \bf f_{\rm h}$. In this case,  
\begin{equation}
\DL \left( \bf\zh, \textbf{r} \right) \approx {1\over 2} \Big|{\delta\over{\bf n}} \Big|^2 + {{\delta}\over{\textbf{n}}}\cdot{{\textbf{r}}\over{\textbf{n}}},
\label{expand2}
\end{equation}
where we have used $\delta\equiv\delta_{\rm m}$ for compactness.
By definition, the noise realization, $\bf r$, is a random variable with zero mean; furthermore,
by construction, the residuals,
$\delta$, are not correlated with $\bf r$. As a result, the second term in equation~(\ref{expand2}) averages to zero:
\begin{equation}
\langle\DL \left( \bf\zh \right)\rangle \approx {1\over 2} \Big|{\delta\over{\bf n}} \Big|^2.
\label{meanDL}
\end{equation}
We can estimate the magnitude of the scatter introduced by the same term by taking the ratio, $\textbf{r}/\textbf{n}$, to be a set of independent normal random variables with unit variance, which results in a standard deviation of
\begin{equation}
{\rm std}(\DL \left( \bf\zh \right)) \approx{} \sqrt{\Big|{\delta\over{\bf n}} \Big|^2} \sim \sqrt{2\langle\DL \left( \bf\zh \right)\rangle}.
\label{stdDL}
\end{equation}
We test this scaling in Appendix~B, where Fig.~\ref{mean:std} shows experiments that highlight the scaling of equation~(\ref{stdDL}).    

In conclusion, from equation~(\ref{meanDL}) we deduce that the mean log-likelihood increase scales with the square 
of the maximum signal-to-noise ratio, $SN_{\rm max}$, and we note that the Bayesian
evidence will also feature in the same scaling. 
From eqns.~(\ref{expandall}) and~(\ref{expand2}) we see that, in real data, 
a scatter of the order of $\sqrt{2\DL }$ should be expected. 
In fact, the same scatter should be expected when mapping the sensitivity function using mock data that include a random noise realization. This implies that multiple noise realizations should be used and results averaged. However, the analysis 
above also shows that this can be avoided by using noise-free mock data (i.e. $\textbf{r}=0$), while at the same time
using the appropriate noise map, $\bf n$, featuring the same maximum signal-to-noise as in the real data. This is 
the strategy we adopt in this work.

\subsection{Fitting procedure}

Having chosen our macromodels, $\tm$, we can introduce intervening LOS haloes with input parameters, ${\zeta}_{\rm h}$, 
and simulate the resulting model fluxes, $\bf{f}(\tm, {\zeta}_{\rm h})=\bf f$.
As outlined in Section~2, each determination of the likelihood gain, $\DL({\zeta}_{\rm h}, \bf{r})$, requires two 
non-linear searches. However, in our case, $\bf r$=0, so that we have, $(\hat\theta_{\rm m}, \hat\zh)=(\tm, \zh)$, 
or equivalently, $\mathscr{L}(\hat\theta_{\rm m}, \hat\zh)=0$, by construction, and therefore, 
\begin{equation}
\DL(\zh) = \mathscr{L}(\bar\theta_{\rm m}) .
\label{DLsimple}
\end{equation}

Thus, for each set of halo parameters, $\zh$, we only require one non-linear search in order to determine the 
best fitting parameters, $\bar\theta_{\rm m}$, of the model that does not include a halo mass component. This is also the fit with fewer free parameters -- and therefore both the fastest to run, and the least likely to get stuck in local minima during a likelihood optimization.

\begin{figure*}
\centering
\includegraphics[width=\textwidth]{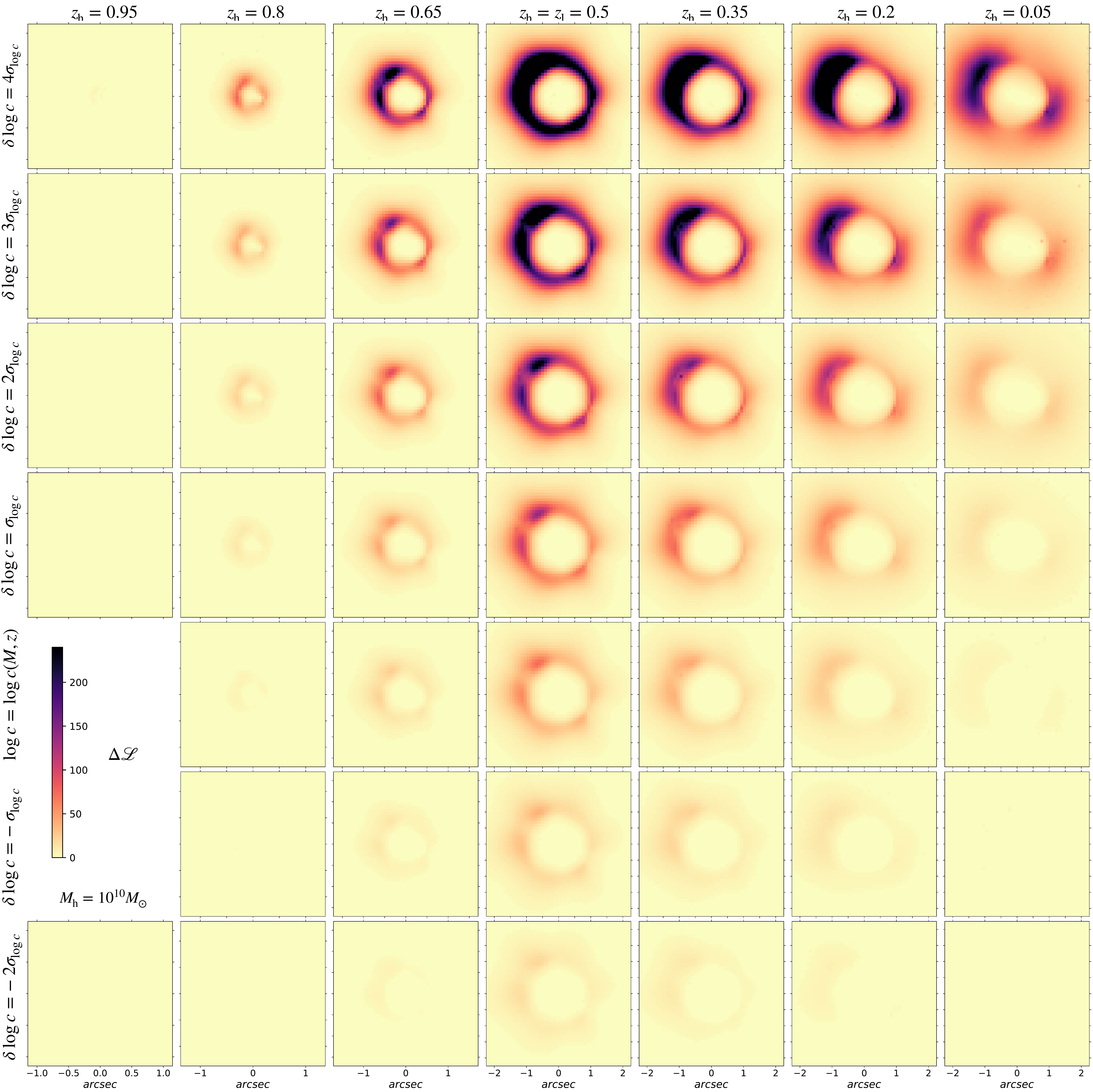}
\caption{An illustration of the sky-projections of our maps of the
  log-likelihood increase, $\DL$, for our `arcs' lensing
  configuration. Columns show a grid of different perturber
  redshifts. Rows are for different perturber concentrations. The
  perturber mass is fixed at $M_{\rm h}=10^{10}$~M$_{\rm \odot}$ in
  all panels. Individual panels share the same colour scale. It is apparent that more concentrated haloes 
result in larger $\DL$ values. Also, the $\DL$ values decrease away from the redshift of the main lens, $z_{\rm l}=0.5$, for both higher and lower perturber redshift.}
\label{arcsmap}
\end{figure*}
\begin{figure*}
\centering
\includegraphics[width=\textwidth]{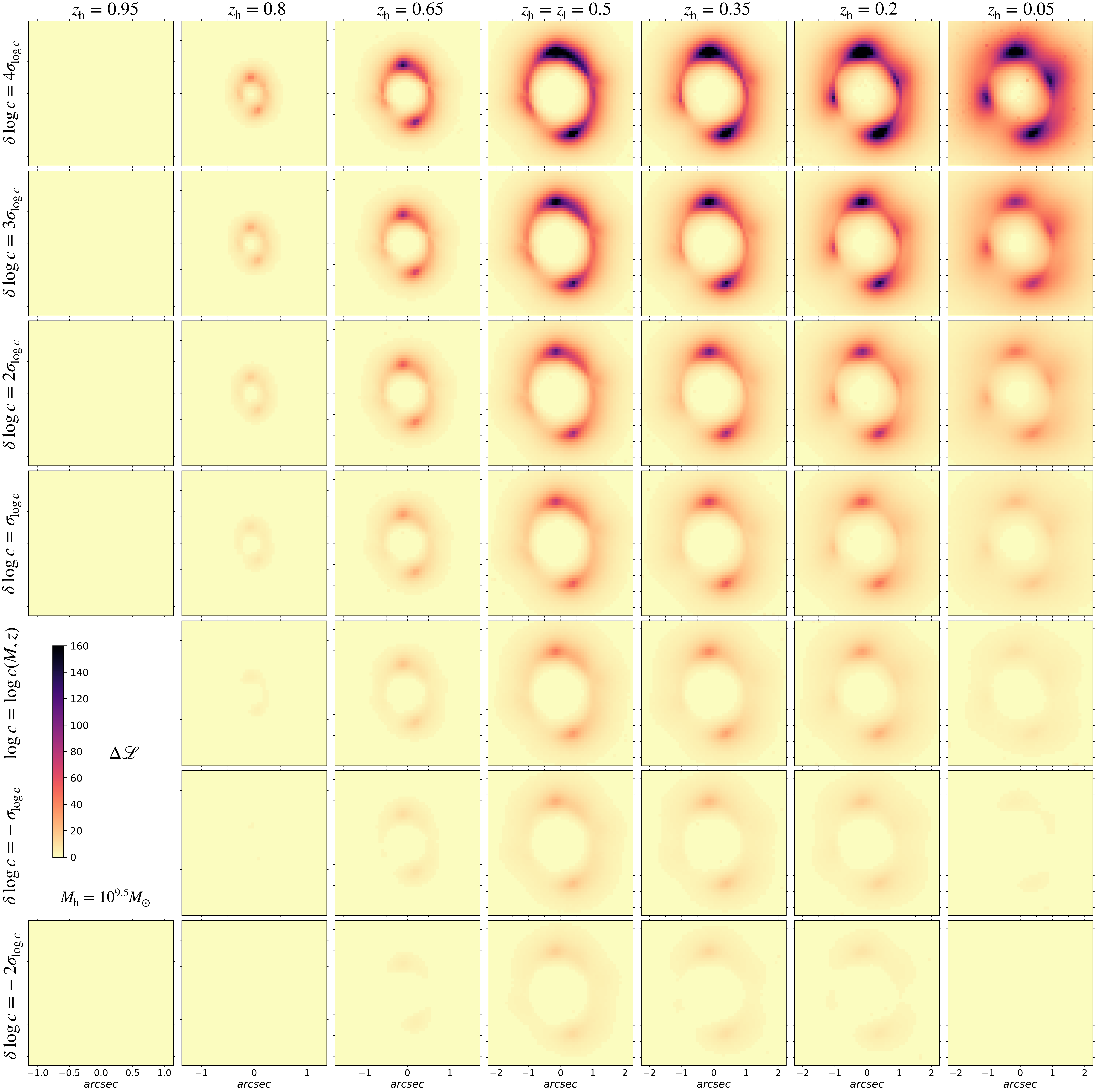}
\caption{Same as Fig.~\ref{arcsmap}, for our `quad' lensing configuration. The perturber halo has a mass of $M_{\rm h}=10^{9.5}$~M$_{\rm \odot}$.}
\label{quadmap}
\end{figure*}

\subsubsection{Gradient descent approach}
We perform these optimizations using an iterative gradient descent
algorithm. In essence, at each step, $i$, provisional  
estimates of the best-fitting parameters, $\tm^i$, and corresponding model fluxes, $\textbf{f}_i$, are used to 
calculate the increments, $\delta\tm^i$, which provide the best linear improvement of the model fluxes themselves.
That is, the increment, $\delta\tm^i$, minimizes the log-likelihood, 
\begin{equation}
\DL_{i+1} ={1\over 2}\sum_{\rm pixels}\Big|{1\over{\bf n}}\Big[\textbf{d}- \Big(\textbf{f}_i+\left.{\partial \textbf{ f}\over\partial \tm}\right|_{\tm^i}\cdot \delta\tm^i \Big)\Big]\Big|^{2},
\label{graddesceq}
\end{equation}
where $\left.{\partial \bf f / \partial \tm}\right|_{\tm^i}$ is the gradient of the model fluxes calculated
at $\tm^i$. This minimization is easily solved by the corresponding least square problem. 
The parameters at the subsequent step are therefore, $\tm^{i+1}=\tm^i+\eta \delta\tm^i$, 
where $\eta$ is the so called learning rate. Iterations are stopped when the corresponding likelihood value converges. 
In order to avoid convergence at possible local maxima, we repeat the procedure over a set of different initialization 
parameters, close to the input parameters, $\tm$. In practice, we find this to be rarely necessary, possibly because
for most perturbing haloes the best-fitting parameters, $\bar\theta_{\rm m}$, are sufficiently close to the input values themselves, and the log-likelihood surface is smooth in our noise-free setting. 

Despite allowing for this redundancy, we find gradient descent to be efficient and inexpensive for models featuring 
parametric sources. This is because the (noise-free) gradient maps, ${\partial \bf f /\partial \tm}$, are well behaved and easily
estimated. In contrast, this is not so when non-parametric pixelized sources are used. We have tried using gradient descent
to optimize the parameters of the lens while, at each iterative step, a linear source inversion \citep[e.g.][]{Warren2003, Dye2005} 
determines the source model. However, we find this approach to be unsuitable, despite the fact that in this case, 
gradient descent is used on a significantly smaller parameter space (featuring 8 dimensions instead of 15). 
Due to the nature of the semi-linear source inversion, the residuals, $\bf d-f_i$, contain 
little information on the mass model itself, unless unrealistically high regularization
values \citep[e.g.][]{Suyu2006} are used.

\section{Mapping the sensitivity function}

For each of the two macromodels we investigate, we map the log-likelihood gain, $\DL$, over the space of halo 
parameters, $\zh=(x_{\rm h}, y_{\rm h}, z_{\rm h}, M_{\rm h}, c_{\rm
  h})$, using a rectangular grid as follows. 
\begin{itemize}
\item{The halo mass is varied between $8.0\leq\log M_{\rm h}/M_{\rm \odot}\leq 10.0$, at intervals of 0.5 dex;}
\item{Halo redshift is varied between $0.05\leq z_{\rm h}\leq0.95$, at intervals of $\delta z = 0.15$. }
\item{Halo concentrations deviate from the mass-concentration relation between 
$\log c - \log c(M,z) \equiv \delta \log c = 4\sigma_{\log c}$ and  $\delta\log c = -2\sigma_{\log c}$, in intervals of $\sigma_{\log c}$. 
Here $\sigma_{\log c}$ is the lognormal scatter of the mass concentration relation, which we take to be independent of 
halo mass and redshift, and fix at $\sigma_{\log c}=0.15$~dex \citep{Wang2020}. 
For this exploration, we assume that $\delta \log c = 0$ means the
mass-concentration relation,  $c(M,z)$, of CDM haloes, 
as measured by \citet{Ludlow2016}.}
\item{Projected locations, $x_{\rm h}, y_{\rm h}$, are mapped over 50 intervals in both coordinates. We scale the total extent of our maps with redshift so as to achieve better
spatial resolution in the $(x,y)$ plane when $z_{\rm h}>z_{\rm l}$.}
\end{itemize}
Figs~\ref{arcsmap} and~\ref{quadmap} illustrate some of our maps as sky-projections of the log-likelihood 
increase, $\DL$, for our `arcs' and `quad' configurations respectively. The first figure shows results for a perturbing halo
of mass, $M_{\rm h}=10^{10}$~M$_{\rm \odot}$, the second for $M_{\rm h}=10^{9.5}$~M$_{\rm \odot}$.
Columns correspond to different values of the perturber redshift, $z_{\rm h}$. Rows are for different halo concentrations.

\begin{figure}
\centering
\includegraphics[width=\columnwidth]{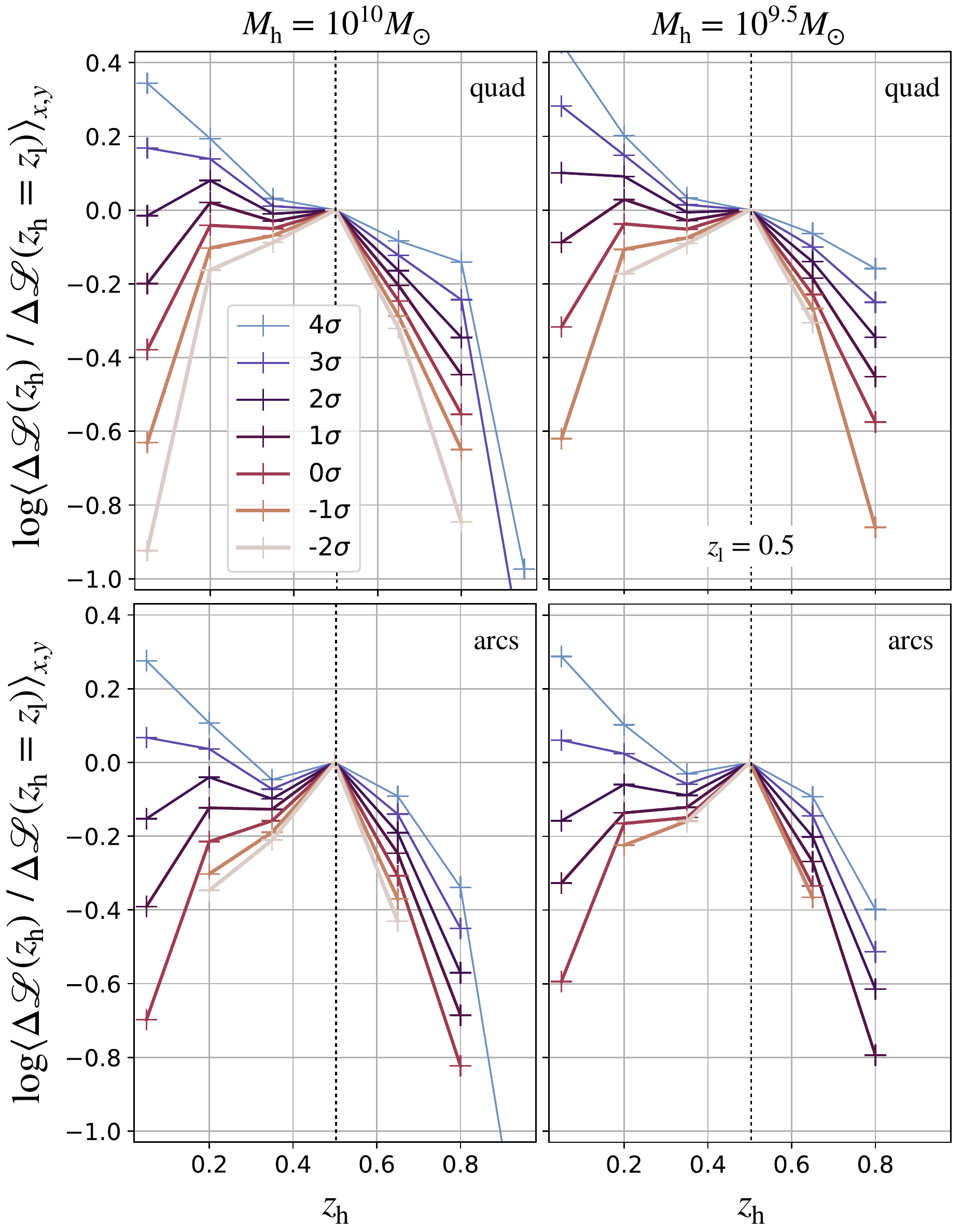}
\caption{The average dependence of the log-likelihood increase, $\DL$, on redshift (see text). The top row refers to our `quad' configuration,
the bottom one to our `arcs'. The right column is for a perturber of
mass, $M_{\rm h}=10^{10}$~M$_{\rm \odot}$, the left column for one of mass $M_{\rm h}=10^{9.5}$~M$_{\rm \odot}$. Profiles of different colours refer to different values of the perturber concentrations (in terms of the shift relative to the median concentration at the relevant redshift; see text). }
\label{redshiftdep}
\end{figure}

\subsection{Dependence on redshift}
\label{redshiftdependence0}

A common assumption in sensitivity mapping is that the perturber's redshift can be recast in terms of its effective mass. 
\citet{Li2016} noticed that the mass and redshift of a perturber are highly 
degenerate, and used this to introduce the idea of rescaling a perturber's 
mass as a function of redshift.
\citet{Despali2018} investigated this equivalence further and provided a universal scaling between redshift and effective mass, which is obtained by requiring that 
the map of deflection angles be minimally changed. In practice, at any fixed projected location, 
the perturbing characteristics of a halo of mass, $M_{\rm h}$, at a redshift, $z_{\rm h}$, have been equated to 
those of a halo located at the redshift of the lens and having an `effective' mass of $\log M_{\rm sh}=\log M_{\rm h}+\delta M(z_{\rm h})$.  
The mass shift, $\delta M(z_{\rm h})$, is clearly zero at $z_{\rm h}=z_{\rm l}$, and is found to be monotonically increasing with 
redshift, so that detecting a halo of fixed mass becomes more challenging with increasing redshift, 
and is easiest for close-by perturbers.

Figs~\ref{arcsmap} and~\ref{quadmap} show that our calculations do not
support this working hypothesis in previous work. When fitting image
fluxes (rather than deflection angles), we find that haloes are harder
to detect when they are behind {\it or} in front of the main lens.  
The details depend on the precise lensing configuration, mass and concentration 
of the perturbing halo, as well as on the precise projected location. However, a decrease in $\DL$ when the halo redshift deviates from the lens redshift 
is a universal qualitative feature in our maps displayed by both our adopted lensing configurations
and across halo masses.
This means that the effect of foreground haloes of fixed mass is more easily `reabsorbed' by suitable 
macromodels when these perturbers are at low redshifts.
In other words, degeneracies in the lens modelling make the detection of perturbers of the same mass
increasingly more difficult with decreasing redshift.  

Fig.~\ref{redshiftdep} summarizes the redshift dependence of the log-likelihood
increase, showing the ratio between the log-likelihood increase for a perturber at some redshift, $\DL(z_{\rm h},M_{\rm h},c_{\rm h})$, 
divided by that at the redshift of the lens,  $\DL(z_{\rm h}=z_{\rm l},M_{\rm h},c_{\rm h})$.
This ratio is then averaged over the perturbers' projected locations, $(x,y)$. The top row refers to our `quad' configuration,
the bottom one to our `arcs'. The right column is for a perturber of
mass, $M_{\rm h}=10^{10}$~M$_{\rm \odot}$, the left column to one of 
mass,  $M_{\rm h}=10^{9.5}$~M$_{\rm \odot}$. Profiles of different colours refer to different values of the perturbers' 
 concentrations. As described, for most haloes, the log-likelihood increase {\it decreases} for redshifts that are
 both higher and lower than the lens' redshift, $z_{\rm l}$. The size of this decrease depends systematically 
 and monotonically on halo concentration, with the less-concentrated haloes displaying sharper falloffs. 

 Depending on the lensing configuration, for haloes on the mass-concentration relation, $\DL(z_{\rm h})$ 
 decreases by a factor between 1.15 to 1.6 between $z_{\rm h}=z_{\rm
   l}$ and $z_{\rm h}=0.2$, and then 
 drops more sharply towards lower redshift.
 For the highest halo concentrations, we see, instead, a mild apparent increase in detectability at low redshift. However, analysis of the top rows 
 of Figs.~\ref{arcsmap} and~\ref{quadmap} shows that this increase is due to the 
 fact that Fig.~\ref{redshiftdep} displays averages over projected
 locations. At the highest concentrations, the projected area in which
 perturbers result in `intermediate' log-likelihood values,
 $50\lessapprox\DL\lessapprox100$ (corresponding to orange hues in
 Figs.~\ref{arcsmap} and~\ref{quadmap}) increases at the lowest redshifts. In the same regions, log-likelihood values are 
 lower at $z_{\rm h}=z_{\rm l}$, which drives the mild increase apparent in the average quantities shown in Fig.~\ref{redshiftdep}. On the other hand, even at the highest concentrations, it remains true that the peak values of the log-likelihood gain, $\DL$, {\it decrease} with decreasing redshift.
 In any case, this mild increase is limited 
 to extremely concentrated haloes, and, therefore, is not a representative behaviour. 
 
 The qualitative contradiction between the predictions using deflection angle maps and our results implies 
 that previous estimates of the number of detectable haloes obtained  using the relation between mass and redshift 
 proposed in \citet{Despali2018} are likely to overestimate the number of low redshift haloes. 
 We will return to this point in Section~\ref{redshiftdependence1}.

\begin{figure}
\centering
\includegraphics[width=\columnwidth]{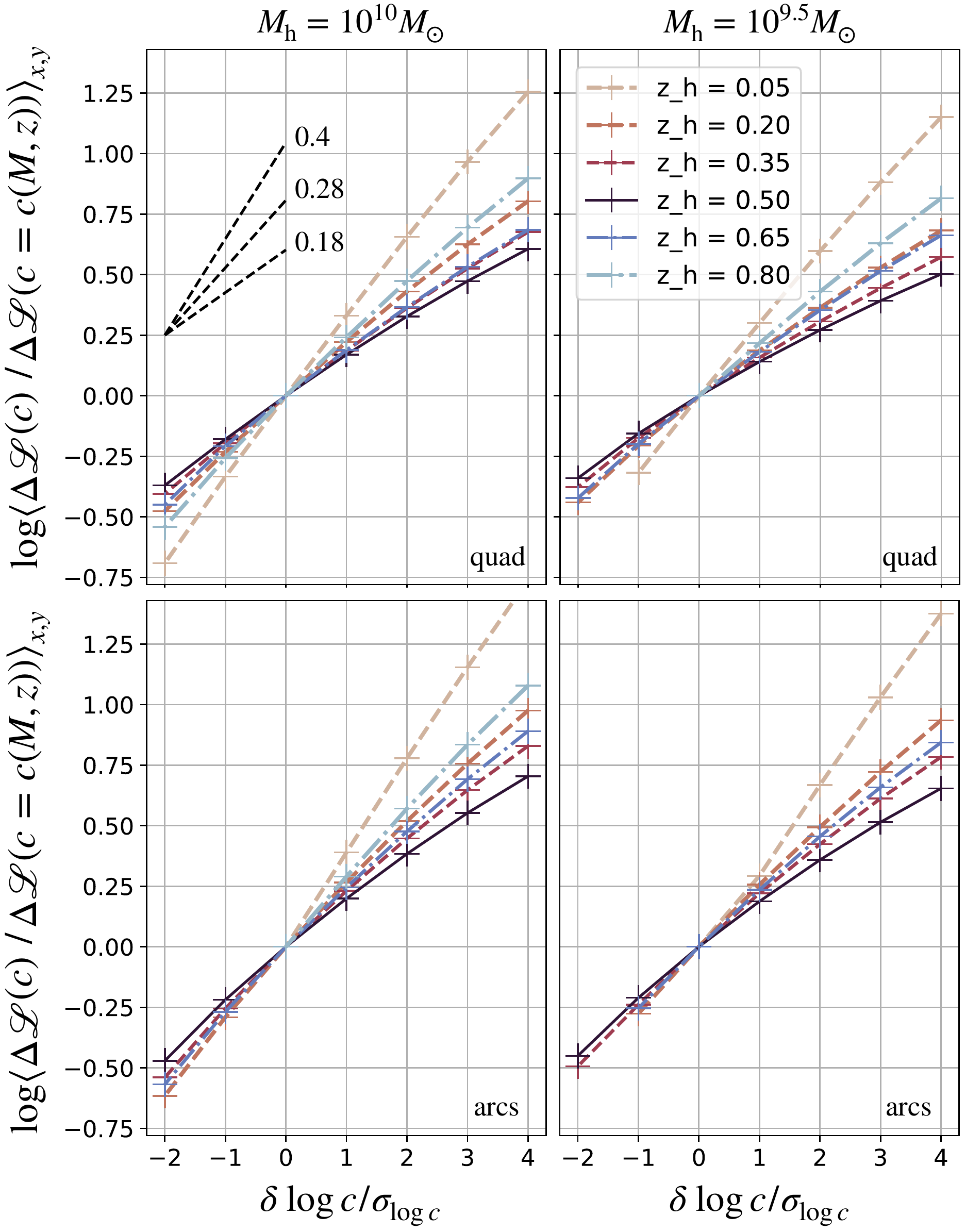}
\caption{The average dependence of the log-likelihood increase, $\DL$, on halo concentration (see text).The top row refers to our `quad' configuration,
the bottom one to our `arcs' configuration. The right column is for a perturber of
mass, $M_{\rm h}=10^{10}$~M$_{\rm \odot}$, the left column to one of mass,
$M_{\rm h}=10^{9.5}$~M$_{\rm \odot}$. Lines of different colours
refer to different values of the perturber  
 redshift. }
\label{concdep}
\end{figure}

\subsection{Dependence on concentration}
\label{concentrationdependence0}

Most previous studies have fixed the halo concentration to 
the mean for their mass for the adopted mass-concentration relation. However, Figs.~\ref{arcsmap} and~\ref{quadmap} clearly show that concentration 
makes a significant difference to halo detectability. From top to bottom, the values of the log-likelihood
gain decrease monotonically: the perturbations of less concentrated haloes are more easily reabsorbed 
by changes of the macromodel parameters. In turn, at fixed mass, more concentrated
haloes are more easily detected. 

Fig.~\ref{concdep} provides a summary of the dependence of the log-likelihood 
increase, $\DL$, on halo concentration. This shows the ratio between the log-likelihood increase,
$\DL(z_{\rm h},M_{\rm h},c_{\rm h})$, for a perturber of any concentration, $c$, divided by that for one on the mass-concentration 
relation, $\DL(z_{\rm h},M_{\rm h},c_{\rm h}=c(M_{\rm h}, z_{\rm h}))$. We reiterate that
$\delta \log c=0$ in our mapping of the sensitivity function corresponds to the mass-concentration relation measured by \citet{Ludlow2016}. Similarly to Fig.~\ref{redshiftdep},
these ratios are then averaged over projected locations, $(x,y)$. The top row refers to our `quad' configuration,
the bottom to one to our `arcs'. The right column is for a perturber
of mass, $M_{\rm h}=10^{10}$~M$_{\rm \odot}$, the left column
for one of mass, $M_{\rm h}=10^{9.5}$~M$_{\rm \odot}$. Profiles of different colours refer to different values of the perturbers' 
 redshift. It is clear that $\DL$ increases monotonically with concentration in all cases. 
 The scalings appear qualitatively similar in all four panels, although, as for the dependence on redshift,
 quantitative details are still dependent on the lensing configuration and other halo parameters. 
 In particular, we record a significant secondary dependence on redshift: the detectability of haloes
 at the lowest redshifts is most strongly boosted by concentration. The magnitude of this boost
 then decreases for redshifts approaching the redshift of the lens, where it has a minimum, to then increase again towards higher redshifts.  
 
 Notably, we find the dependence on concentration to be essentially exponential, al least when averaged 
 over projected locations:
 \begin{equation}
\langle\DL(\delta_{{\rm log}c})\rangle_{(x,y)} \sim 10^{\alpha\cdot \delta_{{\rm log}c}}\langle\DL(\delta_{{\rm log}c}=0)\rangle_{(x,y)} \ .
\label{concexp}
\end{equation}
The top-right panel of Fig.~\ref{concdep} displays guiding lines for the exponent $\alpha=\{0.18,0.28,0.4\}$. 
The log-likelihood increase grows roughly by a factor between 1.5 to 2.5 for each additional $+1\sigma$ 
deviation from the mass-concentration relation. We will analyse the consequences of this on the 
expected number of haloes in Section~\ref{concentrationdependence1}.

\begin{figure}
\centering
\includegraphics[width=\columnwidth]{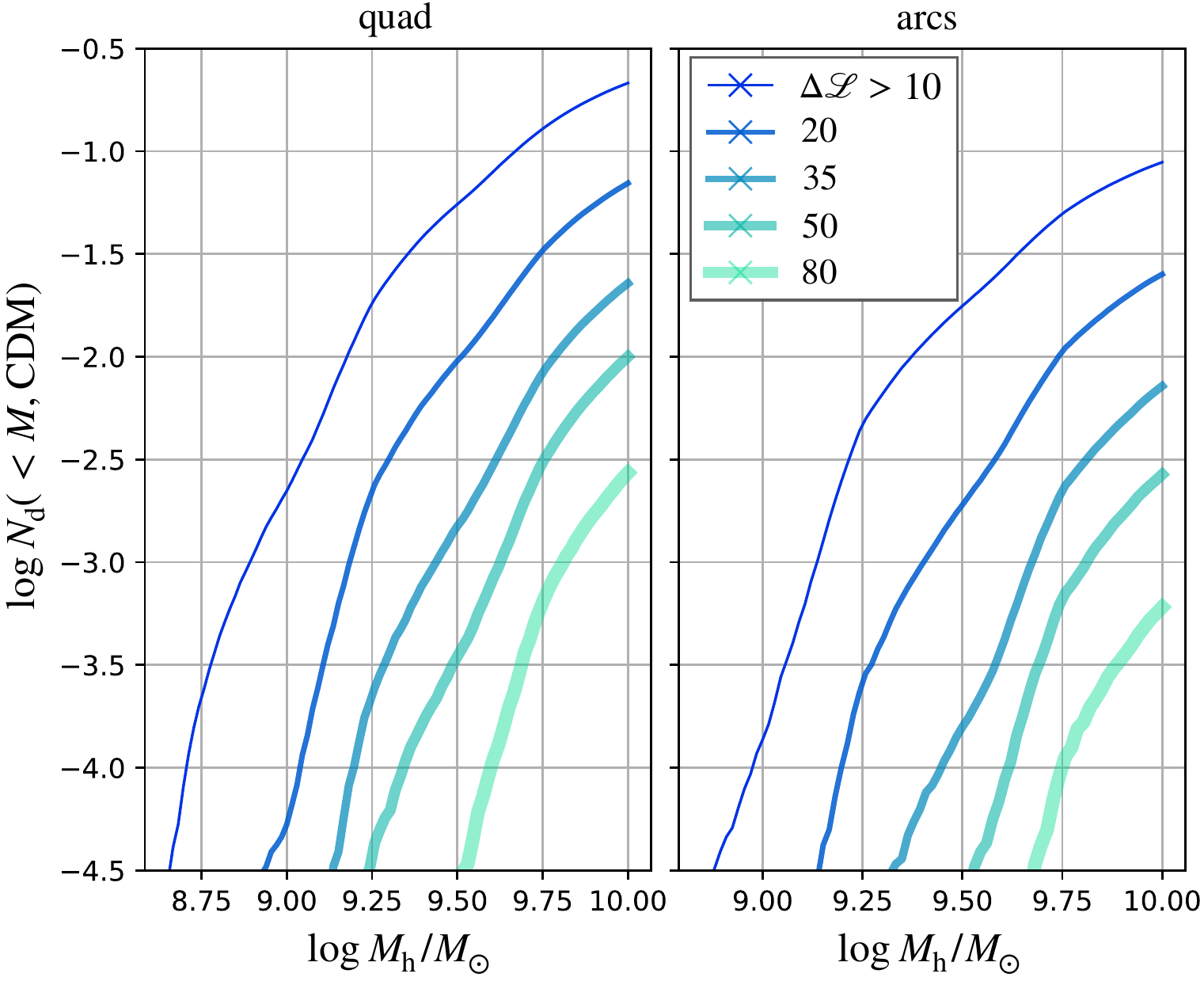}
\caption{The cumulative number of detectable haloes, $N_{\rm d}(<M)$, in a CDM universe. Lines of different colour refer to different
values of the log-likelihood threshold required for detectability.
The left panel refers to our quad configuration, the right panel 
to our configuration featuring asymmetric arcs.  The values displayed 
 are for 
$SN_{\rm max}=50$}
\label{NCDM}
\end{figure}
\begin{figure*}
\centering
\includegraphics[width=.9\textwidth]{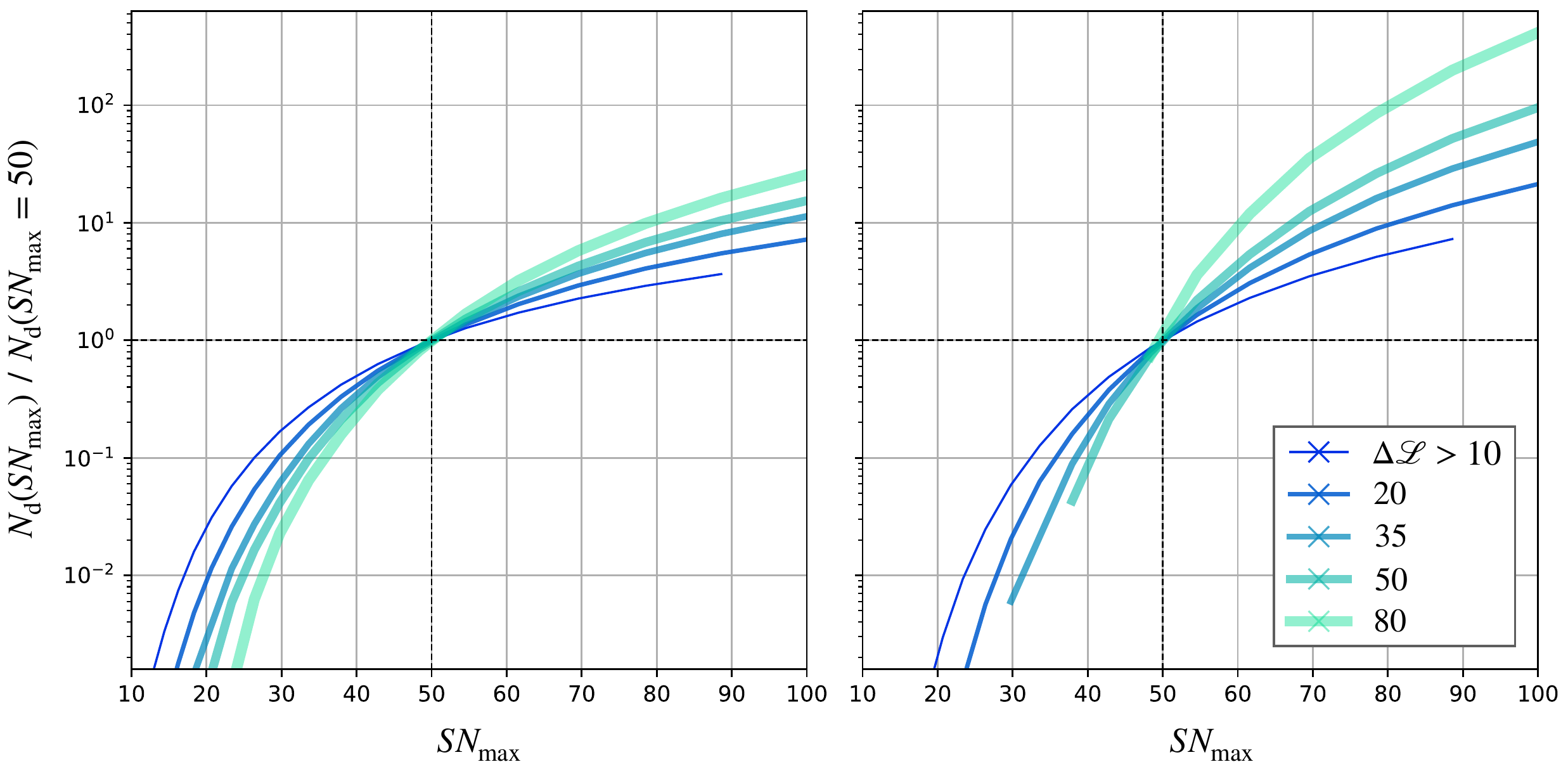}
\caption{The number of detectable haloes of mass, $M_{\rm h}<10^{10}$~M$_{\rm \odot}$
(left panel), and $M_{\rm h}<10^{9.5}M_{\rm \odot}$ (right panel) in a CDM universe as 
a function of the maximum $SN$ of the data. Values are normalised to the number of
detections predicted for $SN_{\rm max}=50$, which we use as a fiducial value in this work. Lines of different colour refer to different
values of the log-likelihood threshold required for detectability.}
\label{Nofsn}
\end{figure*}

\section{The population of detectable haloes}

Using our maps of the log-likelihood increase, $\DL(\zh)$, we are ready to perform the integral in 
eqn.~(\ref{ndetectable}). As done in previous work, we use a sharp threshold, $\DL_{\rm th}$, in the 
log-likelihood increase to separate detectable haloes from non-detectable haloes:
 \begin{equation}
 p(\zh) = 
 	\begin{cases}
	1 &{\rm if}\  \DL(\zh)\geq \DL_{\rm th}\\
	0 &{\rm if}\  \DL(\zh)< \DL_{\rm th},\\
	 \end{cases}
\label{pofdl}
\end{equation}
although we notice that the scatter characterized in equation~(\ref{stdDL}) would provide a 
{natural} scaling for a smooth transition in the detection probability. This will be useful when preparing detailed 
predictions for real data, but would not affect our conclusions here,
so in this study we retain a sharp transition for simplicity.


We parametrize the cosmological number density of DM haloes as suggested by \citet{Lovell2014}:
 \begin{equation}
n(M_{\rm h},z | M_{\rm cut}) = n_{\rm CDM}(M_{\rm h},z)\Big(1+{M_{\rm cut}\over M_{\rm h}}\Big)^{-1.3},
\label{halomassf}
\end{equation}
where $n_{\rm CDM}(M_{\rm h},z)$ is the CDM halo number density, for which we adopt the form
derived by \citet{Sheth2001}. We take this distribution to be uniform over projected sky coordinates,
and assume that the distribution in concentration is lognormal, with a spread of $\sigma_{\log c}=0.15$~dex 
(independent of mass, redshift and DM model). We take the median concentration to be dependent on
the DM model, and adopt the parametrization proposed by \citet{Bose16}:
\begin{align}
&c(M_{\rm h},z | M_{\rm cut}) = \nonumber  \\
&c_{\rm CDM}(M_{\rm h},z)\Big[(1+z)^{0.026z-0.04}\Big(1+60{M_{\rm cut}\over M_{\rm h}}\Big)^{-0.17}\Big],
\label{mcrel}
\end{align}
in which the median concentration of CDM haloes is as recorded by \citet{Ludlow2016}.

These prescriptions allow us to calculate the integral of equation~\ref{ndetectable} using a Monte Carlo strategy.
We randomly sample the candidate haloes' $\zh$ according to their cosmological number density, and then check 
whether they would be detectable using our maps of log-likelihood increase, which we linearly interpolate 
between our rectangular grid points.

\subsection{The effect of data quality}

Although our objective is not to provide absolute figures for the 
number of detectable haloes, Fig.~\ref{NCDM} shows the cumulative distributions
of detectable LOS haloes, $N_{\rm d}$, we obtain for our two lens configurations. Both panels
are for a CDM universe and are calculated from our full maps of the 
log-likelihood increase, that is including both the full redshift dependence 
and the scatter in the mass concentration relation. 
We have used $SN_{\rm max}=50$. We stress that these figures cannot be directly applied
to real data analysed with different techniques and featuring different
lensing configurations. For definitiveness, we include all haloes with projected
coordinates in a $4.5''\times4.5''$ area at $z_{\rm h}\leq z_{\rm l}$, decreasing
to $2.1''\times2.1''$ at $z_{\rm h}=z_{\rm s}=1$, as displayed in Figs.~\ref{arcsmap} and~\ref{quadmap}.

Our two lensing configurations provide substantially different 
numbers of detectable haloes: a quad configuration appears less prone
to modelling degeneracies and therefore more promising for the detection
of perturbers. The number of expected detections is also a strong function of
the imposed detection threshold. 
In our quad lens configuration, thresholds of $\DL_{\rm th}=\{$10, 20, 35, 50$\}$ yield $N_{\rm d}=\{$0.2, 0.07, 0.02, 0.01$\}$ detections with $M_{\rm h}<10^{10}$~M$_{\rm \odot}$ 
per lens, respectively. 

The analysis of Section~\ref{stonoise} also allows us to address systematically
the dependence of the total number of detected haloes with the quality
of the 
data, which in this work we have characterized by the value of 
$SN_{\rm max}$. Equation~\ref{meanDL}
allows us to equate changes in the value of $SN_{\rm max}$ with 
changes in the value of the log-likelihood threshold required for 
detection, $\DL_{\rm th}$: 
\begin{equation}
N_{\rm d}(SN_{\rm max}, \DL_{\rm th}) = N_{\rm d}(\alpha SN_{\rm max}, \alpha^{-2}\DL_{\rm th})\ ,
\label{snchange}
\end{equation}
for any factor $\alpha$. 
We use this equivalence to focus on how 
the number of expected detections for a CDM universe would change
for higher or lower values of the signal-to-noise, i.e.\ for longer or shorter 
exposure times.

Fig.~\ref{Nofsn} shows the number of expected detections
of haloes of mass, $M_{\rm h}<10^{10}$~M$_{\rm \odot}$
(left panel) and $M_{\rm h}<10^{9.5}$~M$_{\rm \odot}$ (right panel) for varying
values of $SN_{\rm max}$, normalized by the number of detections
predicted for $SN_{\rm max}=50$. The figure displays the case of our
`quad' configuration; results for our `arcs' configuration are similar, 
albeit with a more marked dependence on the $SN$ itself.
Lines of different colour refer to different
values of the log-likelihood threshold required for detectability.
As expected, the number of detectable haloes is a rapidly increasing function
of the signal-to-noise ratio. We find that, for a likelihood threshold 
of $\DL_{\rm th}=20$, an increase in signal-to-noise ratio from 
50 to 60 corresponds to a doubling in the number of expected detections
$N_{\rm d}(M_{\rm h}<10^{10}$~M$_{\rm \odot})$. The same increase in data quality 
makes the expected detections $N_{\rm d}(M_{\rm h}<10^{9.5}$~M$_{\rm \odot})$
grow by a factor of three. As shown in the same figure, these factors 
are even larger if higher values of the log-likelihood 
ratio are required for detectability. For a value of $\DL_{\rm th}=50$,
analogous to what has been used in most previous studies, we find the corresponding
figures to be 2.5 and 5 for haloes of $M_{\rm h}<10^{10}$~M$_{\rm \odot}$ and 
$M_{\rm h}<10^{9.5}$~M$_{\rm \odot}$ respectively. It should be noted that
an increase in the maximum $SN$ ratio from 50 to 60 corresponds to 
an increase in the exposure time of a factor of $\approx 1.44$, {which is therefore smaller than the corresponding gain
in the number of detectable haloes.}

\begin{figure}
\centering
\includegraphics[width=.8\columnwidth]{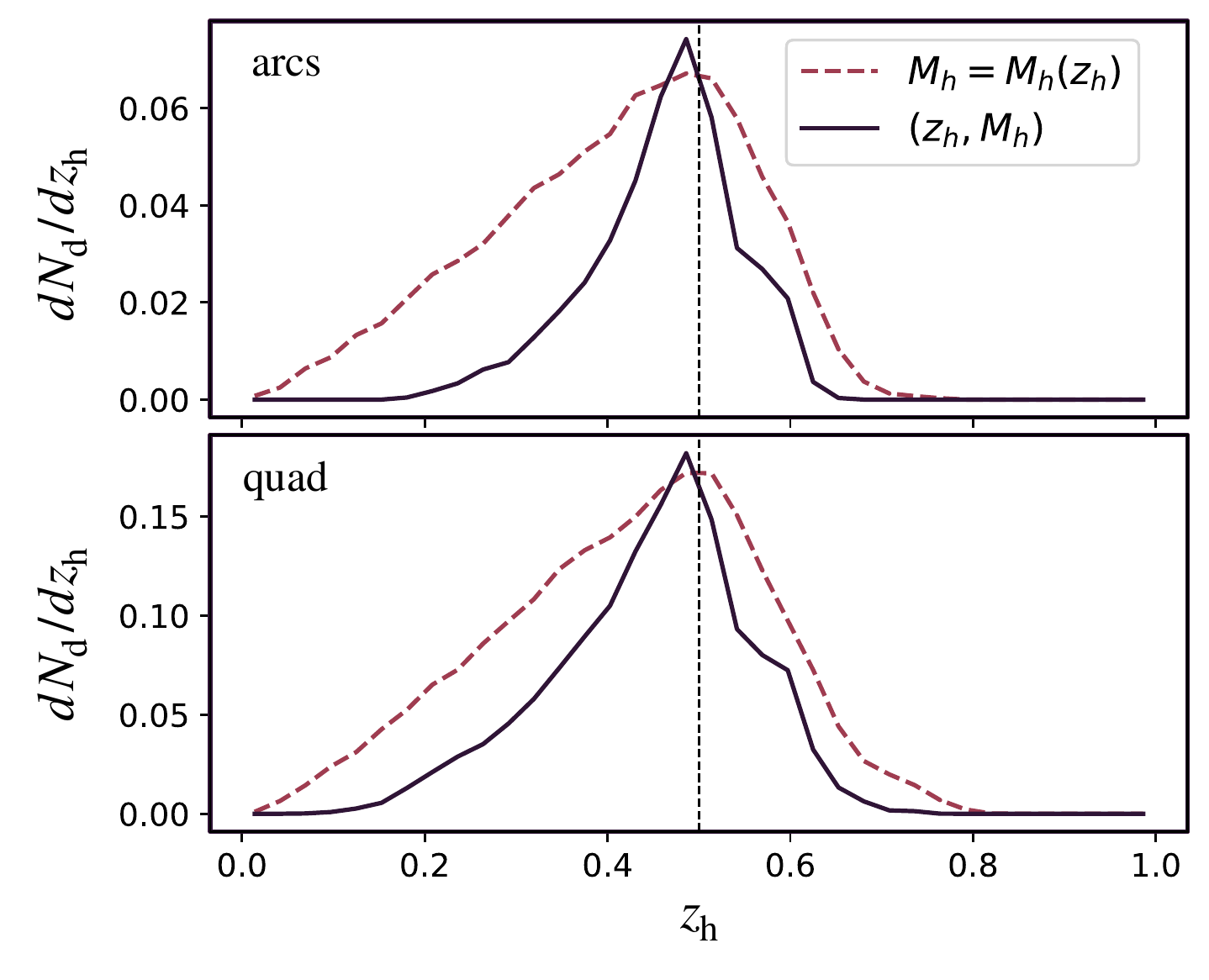}
\caption{Predictions for the redshift distribution of detected CDM haloes obtained when:
(i)~using the relation between halo mass and redshift proposed by \citet{Despali2018} (dashed line),
(ii)~using the full redshift dependence of the log-likelihood increase
(solid line). The top panel is for the arcs configuration, the
bottom panel for the quad configuration. Concentration effects
are not included in this comparison.}
\label{nredshift}
\end{figure}

\subsection{The effect of redshift dependence}
\label{redshiftdependence1}

We now examine the consequences of dropping the simplifying assumption
of a tight relationship between halo mass and redshift. 
We isolate this effect by considering a population of CDM haloes
assumed to lie on the \citet{Ludlow2016} mass-concentration relation,
i.e.\ with reference to the previous Section, here we ignore the
scatter in halo concentration. (We will consider this shortly.) Fig.~\ref{nredshift} shows 
the redshift distribution of the population of detected haloes of
mass, $M_{\rm h}<10^{10}$~M$_{\rm \odot}$, we obtain when:
for our two lensing configurations
\begin{itemize}
\item{using the mass shift proposed by \citet{Despali2018} and described in Section.~\ref{redshiftdependence0}, 
shown by a dashed line\footnote{We use the same 
mass shift, $\delta M(z)$, for all projected coordinates, $(x,y)$.};}
\item{using the full redshift dependence of our log-likelihood maps, shown by a solid line.}
\end{itemize}
As expected, the two curves match at $z_{\rm h}=z_{\rm l}$, but we
predict significantly fewer detections 
for foreground haloes, a reflection of the dependence on redshift of the log-likelihood increase described in 
Sect.~\ref{redshiftdependence0}. We also find that collapsing the
redshift axis leads to an overestimate of 
detectable haloes also at $z_{\rm h}>z_{\rm l}$, though by a smaller factor. 

The magnitude of the global overestimate varies with the lensing
configuration.  For definitiveness, we use a threshold of 
$\DL_{\rm th}=20$ in Fig.~\ref{nredshift}. For the arcs configuration, 
the overestimate is  a factor 1.95; for the quad configuration it is
a factor of 1.63. 
We stress that these figures should not be used to `correct' previous measurements of the number of detectable haloes 
and are meant only as an estimate of the magnitude of the effect.

\begin{figure}
\centering
\includegraphics[width=.9\columnwidth]{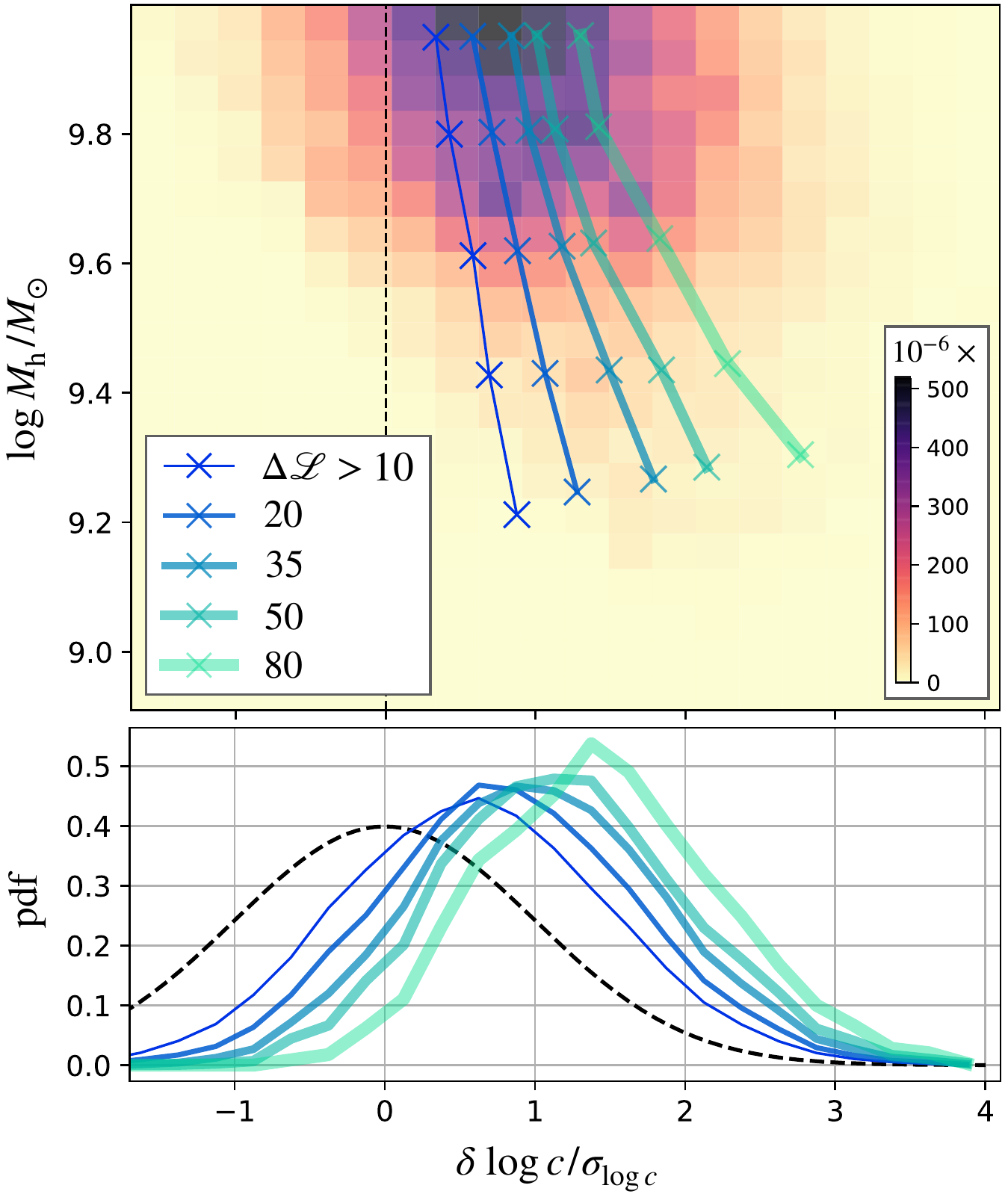}
\caption{The distribution of halo concentration for the population of detectable haloes in a CDM universe.
The top panel shows a 2 dimensional histogram of all detectable haloes in the plane of halo mass
and concentration shift from the median relation, in units of the lognormal spread in the 
mass-concentration relation. The colour scale shows the number of 
perturbers in each M-c pixel for a detection threshold of 
$\DL_{\rm th}=35$. Lines of different colours display the mean concentration shift, $\log c/\sigma_{\log c}$, as a function
of mass for different thresholds for detectability (see text). 
The bottom panel shows the distribution of concentration for all
detectable haloes of $M_{\rm h}<10^{10}$~M$_{\rm \odot}$, as a function of the thresholds for detectability. For reference, the dashed line shows the 
parent distribution of all cosmological haloes. }
\label{cshift}
\end{figure}
\begin{figure}
\centering
\includegraphics[width=\columnwidth]{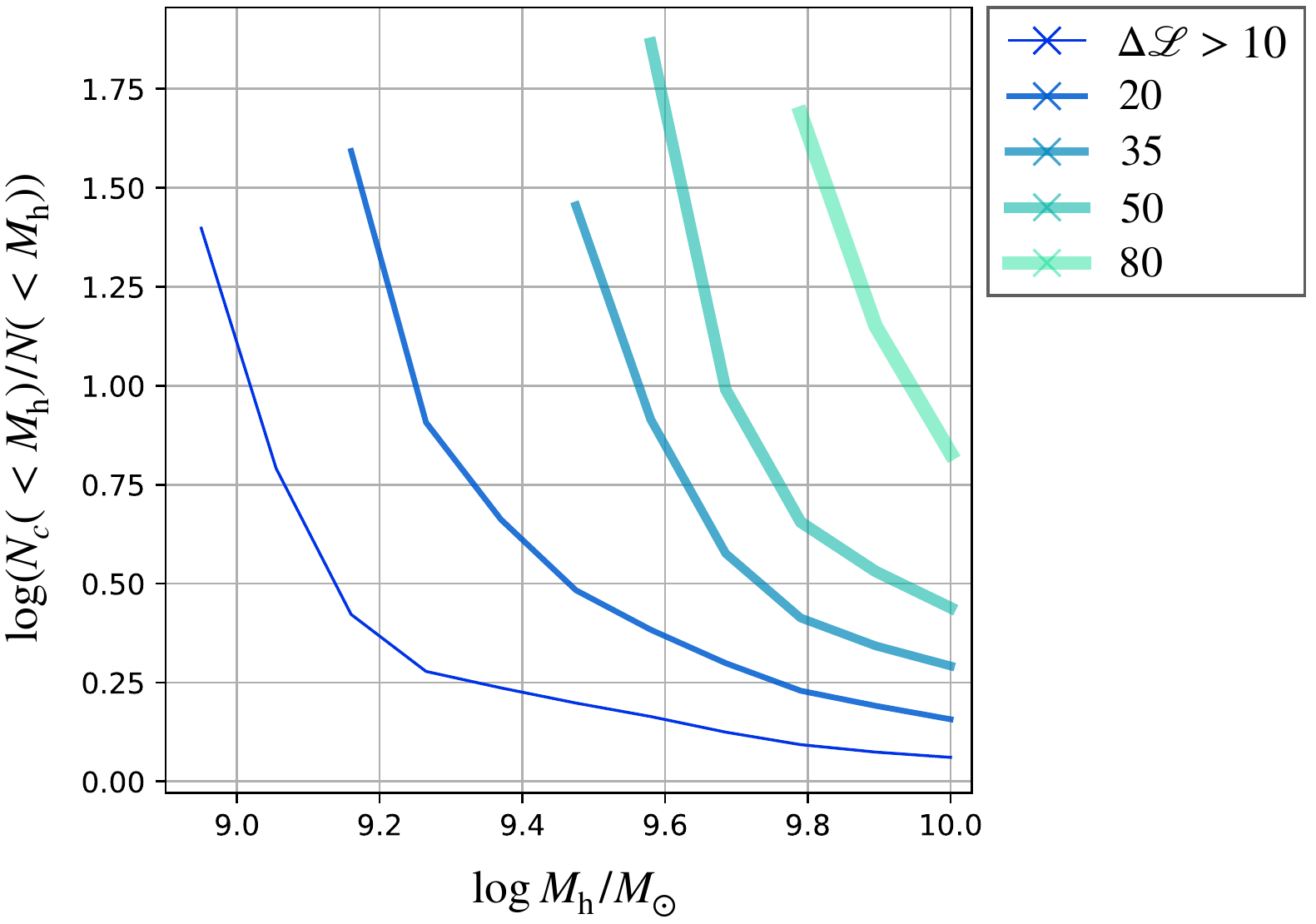}
\caption{The boost to the cumulative number of expected detections, $N_c(<M_{\rm h})$, resulting 
from  including the scatter in the mass-concentration relation for our `quad' configuration. 
Lines of different colours correspond to different thresholds for detectability. }
\label{Nshift}
\end{figure}
\begin{figure}
\centering
\includegraphics[width=.8\columnwidth]{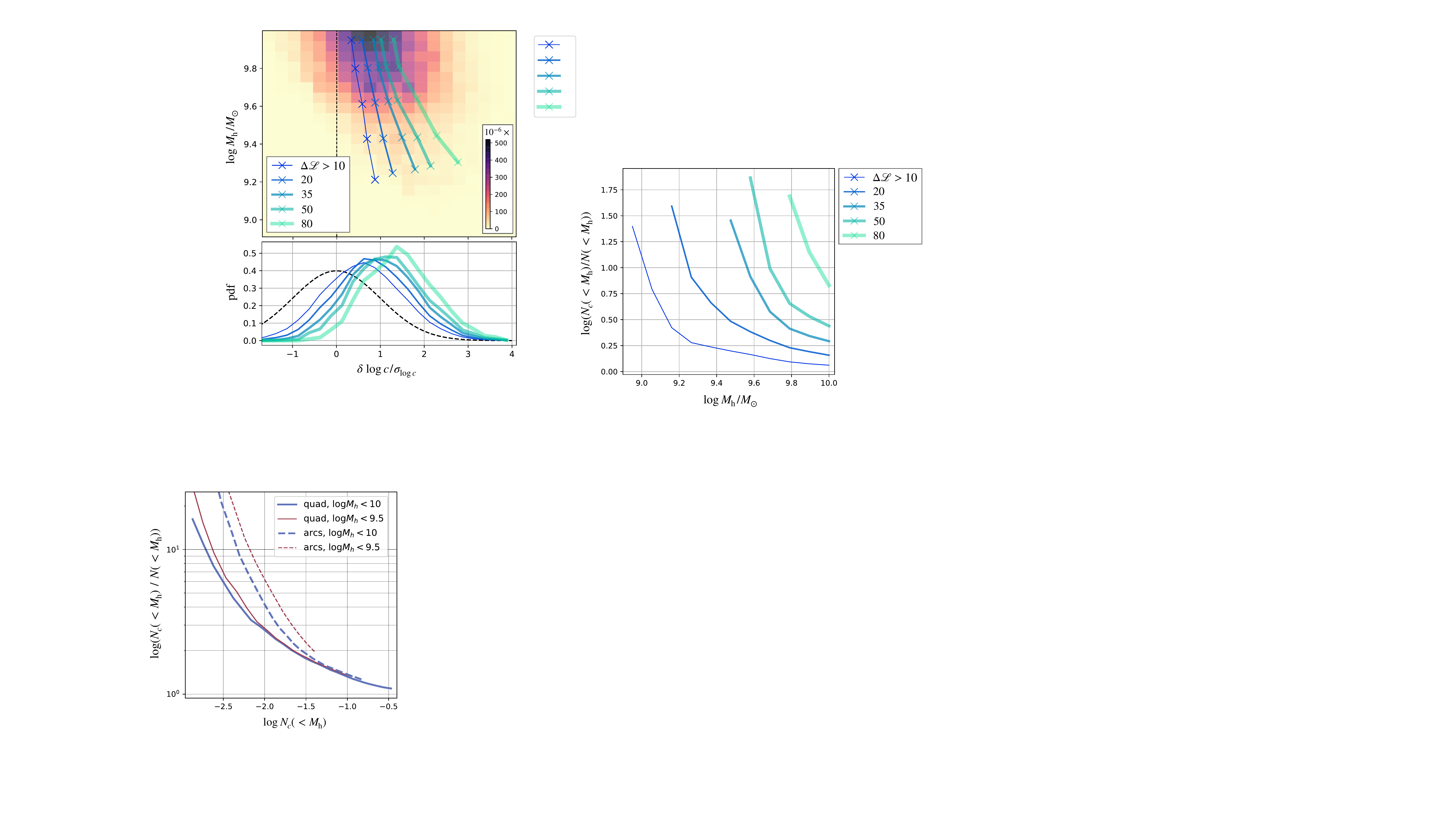}
\caption{The boost to the cumulative number of expected detections, $N_c(<M_{\rm h})$, resulting 
  from including the scatter in the mass-concentration relation, as a function of the 
expected number of detections.}
\label{Ncofn}
\end{figure}

\subsection{The effect of the scatter in concentration}
\label{concentrationdependence1}

We now consider the effect of accounting for scatter in the mass-concentration relation. 
We again focus on a population of CDM haloes, and compare the case in which all haloes 
are assumed to lie exactly on the median mass-concentration relation, to the case in which a lognormal 
scatter is included. For definitiveness, in both cases we use the full redshift dependence of the 
log-likelihood increase, $\DL$, and, for simplicity, we restrict
attention to our quad configuration. 
Results are analogous for our `arcs' lensing morphology.

Fig.~\ref{cshift} shows the distribution of detectable haloes in the space of halo mass
and shift relative to the median halo concentration, for the case in which the scatter in concentration is accounted for. 
The 2 dimensional histogram in the top panel is for an assumed
threshold, $\DL_{\rm th}=35$. The vertical 
dashed line shows the median mass-concentration relation, $\delta\log c=0$. It is clear that,
thanks to the dependence on concentration of the log-likelihood increase described in 
Section ~\ref{concentrationdependence0}, most detectable haloes are high-concentration haloes. 
This is quantified in the bottom panel of the figure, which collapses the mass axis to show 
the distribution of detected haloes over concentration shifts. For reference, the dashed line 
shows the Gaussian distribution of all cosmological haloes. Coloured lines show the distribution of
the population of detectable haloes for different values of the
log-likelihood threshold,  $\DL_{\rm th}$. 

Haloes with high concentration achieve $\DL>\DL_{\rm th}$ more easily, so that higher
thresholds for detectability correspond to increasingly concentrated populations of
detectable haloes. For $\DL_{\rm th}=50$ and our quad configuration, we find
$\langle\delta\log c/\sigma_{\log c}\rangle = 1.25$ when including all
haloes of $M_{\rm h}<10^{10}$~M$_{\rm \odot}$.   
The average concentration shift increases further for decreasing
perturber masses, 
as shown by the coloured lines in the top panel. These represent the `mass-concentration relation
of detectable haloes'. Lower mass haloes require stronger concentration boosts to achieve 
detectability, so that for $\DL_{\rm th}=50$, 
$\langle\delta\log c/\sigma{\log c}\rangle = 2.2$ for haloes of $M_{\rm h}\approx10^{9.2}$~M$_{\rm \odot}$.

\begin{figure*}
\centering
\includegraphics[width=.9\textwidth]{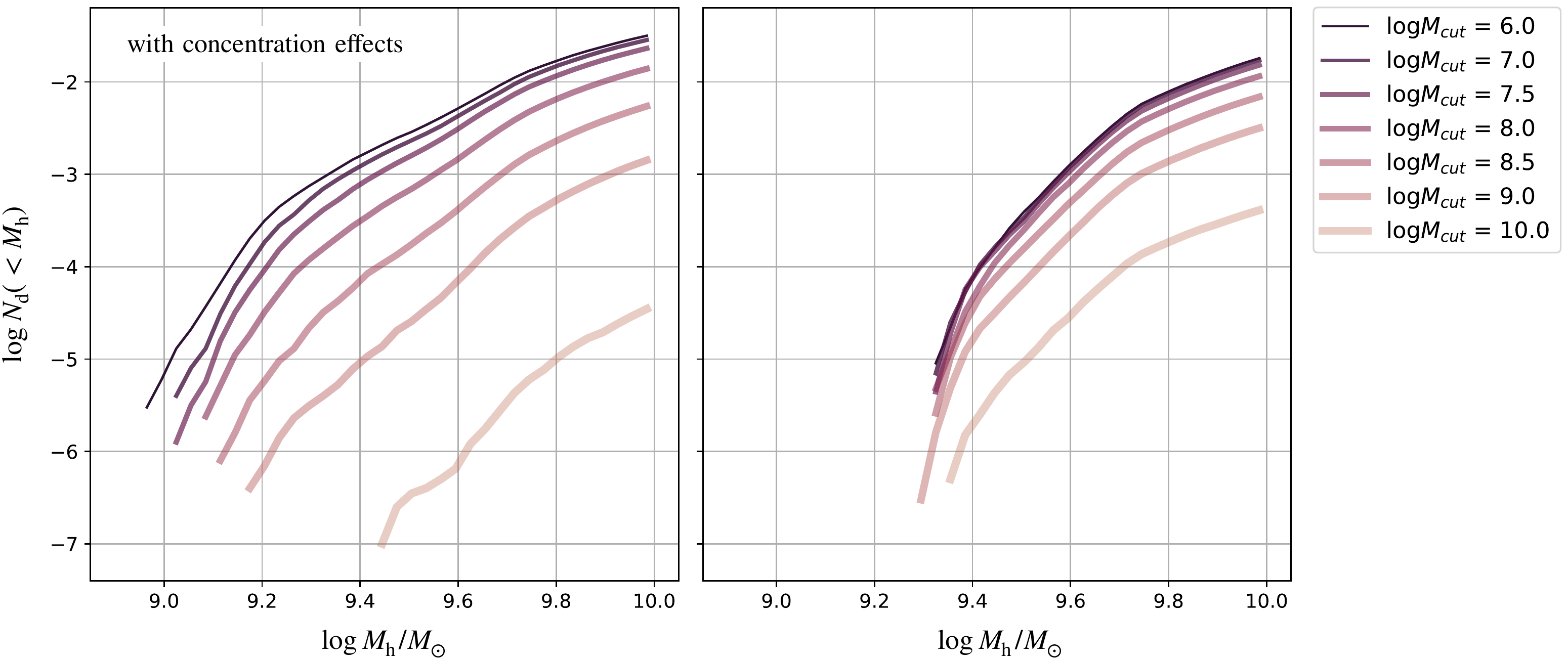}
\caption{A comparison between the cumulative number of expected detections for our `quad'
configuration (threshold for detection, $\DL_{\rm th}=30$) when including both 
(i)~the scatter in the mass-concentration relation and (ii)~the 
dependence of the median concentration on the DM model (left), and when  
concentration effects are neglected (right). Concentration substantially 
enhances the spread between
the expected detections in WDM models with different cutoff masses, $M_{\rm cut}$.}
\label{Nofmhf}
\end{figure*}

The result is that the scatter in the mass-concentration relation
boosts the number of expected detections. {At low halo masses, for any fixed halo mass,} there is a fraction of
haloes with high enough concentration that become detectable if the scatter in the mass-concentration
relation is accounted for, and which is lost when $c=c(M,z)$.  
This is quantified in Fig.~\ref{Nshift}, which shows the ratio between the cumulative number of 
detectable haloes, $N_c(<M_{\rm h})$, predicted when accounting for the scatter in the mass-concentration 
relation, to the corresponding number obtained when $c=c(M,z)$ for all haloes. 
Lines of different colours refer to different values of the detection
threshold, $\DL_{\rm th}$.
For all thresholds, the ratios are a strong function of halo mass. The
onset of the sharp rise identifies
the halo mass that is not detectable at that threshold if $c=c(M,z)$, but for which detections
are possible because of concentration effects. For $\DL_{\rm th}=50$, even including all haloes with $M_{\rm h}<10^{10}$~M$_{\rm \odot}$,
concentration effects boost the number of detectable haloes by a factor of 2.75 for our `quad'
configuration. Since our `arcs' configuration leads to lower values of the log-likelihood
increase and fewer detections, the corresponding boost is a factor $\approx26$, 
exemplifying, on the one hand, the importance of concentration effects, 
and, on the other, the need for estimates tailored to the specific lensing configuration. 

Fig.~\ref{Ncofn} displays the same boost factor as a function of the number of expected
detections, $N_c$ (which include the effect of
concentration). Different values of the expected number of detections correspond
to different values of the data $SN$ (or equivalently, different values of the log-likelihood threshold for detection). Once
again we see that the exact figures depend on the lens configuration;
for example, 
here the arcs configuration appears more sensitive to concentration
effects than the quad.

\subsection{The effect of concentration on distinguishing DM models}
\label{concentrationdependence2}

In most previous estimates of the dependence of the number of
detectable haloes on the properties of the DM model, particularly the
mass of a WDM particle (or, equivalently, the cutoff mass in the
mass function) it was assumed that all haloes lie exactly on the
median mass-concentration relation of CDM haloes.
Not only is this a poor approximation but it left the differences in
the halo abundances alone as the statistic to differentiate amongst
different models. Our results show that concentration is a crucial
ingredient for halo detectability.  Warmer WDM models make low-mass
haloes that are progressively less concentrated, as quantified by
equation~\ref{mcrel}. This makes the concentration parameter potentially helpful in
boosting the spread among the expected numbers of detections in
different DM models.

We test this in Fig.~\ref{Nofmhf}, in which we show, side by side, the cumulative number
of expected detections, $N_{\rm d}(<M_{\rm h})$, for a range of WDM models,
parameterized by their cutoff mass, $M_{\rm cut}$. The panel on the right displays results
for the case in which concentration effects are neglected whereas the
panel on the left includes 
both: (i)~the scatter in the mass-concentration relation and (ii)~the dependence of the median concentration
on the DM model. For definitiveness, we show results pertaining to our `quad' configuration,  
using a threshold for detection of  $\DL_{\rm th}=30$. 

It is clear that concentration effects significantly enhance the dependence of the 
expected number of detections on the DM model. Reduced 
concentration values act together with 
reduced cosmological abundances to determine the expected number of detections. 
Once again, while the qualitative trend is clear and the magnitude of the effect significant, 
precise values are dependent on the lensing configuration and on the value of the 
threshold chosen for detection. We attempt to quantify how much concentration 
effects can actually sharpen substructure lensing constraints in Section~\ref{final}.

\section{Discussion and Conclusions}

We have quantified the ability of low-mass DM haloes along the
line-of-sight to perturb strong gravitational lenses, and explored how
this depends on halo properties.  This is a fundamental ingredient of
sensitivity mapping, that is the process of assessing which
perturbers, out of the cosmological population of haloes, would
actually be detectable when modelling strong lensing data.  It is
impossible to quantify the number of expected detections in different
DM models in a given observational dataset without building the 
sensitivity function. Therefore, sensitivity mapping is a key aspect
of placing constraints on the identity and properties of DM from the
number of perturbing haloes detected in strong lensing studies.

We have adopted a likelihood-based approach, i.e.\ we differentiate
between detectable and non-detectable haloes according to the likelihood
gain, $\DL$, associated with including a halo mass component in the
lens modelling. Some previous studies have proposed using instead the Bayesian
evidence for this comparison. It should be stressed that
both approaches are equally valid. As long as the same criterion is
consistently applied to both measure the sensitivity function — and
therefore make predictions for the different DM models — and detect
 perturbers in the data, both approaches will return the correct inference on the DM properties {if the models used in sensitivity mapping 
 share the same complexities of the real data}. 
 
At this stage, it cannot be excluded that the sensitivity functions
derived using likelihood or Bayesian evidence may exhibit some
differences when compared side by side, possibly reflecting that the
two criteria may lead to different sensitivity to  perturbers
in different regions of the parameter space. However, it seems quite
unlikely that this would happen systematically at the high levels of
significance that we are considering here and that have become the
norm in structure lensing. At present, an evidence-based sensitivity function
has not been derived as a function of either redshift or halo
concentration, where our major findings are. Therefore, we are unable
to make a direct comparison and have  limited our analysis to the differences with the approximate strategies that have been adopted so far.

Rather than attempting to quantify the number of expected detections for 
specific observed strong lenses, we have focused on building an understanding
of the sensitivity function itself, and how this scales with some of the crucial 
parameters at play. We have concentrated on the importance of image 
depth, and 
shown that, as the log-likelihood difference 
scales with $SN^2$, high signal-to-noise data are extremely beneficial in that they allow
the detection of a larger number of perturbers, particularly of low-mass 
haloes (see Fig.~\ref{Nofsn}). We have also shown that the specific noise realization introduces
scatter in the log-likelihood difference, of magnitude $\approx\sqrt{2\DL}$ 
(see Sect.~\ref{stonoise}), 
which suggests that a smooth link
between the probability of detection, $p$, and the log-likelihood
gain, $\DL,$ rather than a sharp threshold, 
may be a better choice {when comparing with} real data. 

We find that our two different lens configurations yield significantly different
numbers of expected halo detections, which indicates that some lensing morphologies
(a quad configuration in our case) are more valuable for strong lensing 
analyses (see Fig.~\ref{NCDM}). This will be useful in selecting
lenses to target with deeper observations.
We also note that our estimates for the total number of detectable haloes are 
somewhat lower than what has been suggested by similar studies for the same values of
lens and source redshifts, and for similar data quality. This may be due to 
the increased flexibility of our lens modelling, including the possibility of
shifts in the lens centre \citep[see][]{Vegetti2014}, as well as the assumed power-law profile slope \citep[see][]{Li2016, Despali2018}. 
With particular reference to \citet{Li2016} and \citet{Despali2018},
the fact that we perform fully non-linear searches when optimising our
macromodels certainly enables  
them to reproduce better the perturbed data without the need of including 
a halo mass component, hence lowering the log-likelihood gain. If anything, 
this highlights the importance of using exactly the same techniques to both i)~model 
real data and make perturber detections and ii)~produce estimates of the expected 
detections, as any mismatch would inevitably introduce systematic biases. 

We then concentrated on the role of halo redshift 
and halo concentration. In previous work, 
simplifications had been made to collapse these axes, in order to make
the calculation of the sensitivity function computationally feasible. 

Concerning the redshift of the perturber, we have shown that, contrary to previous understanding,
it becomes increasingly challenging to detect perturbing haloes in front of the main lens 
when they get closer to the observer (see Figs.~\ref{arcsmap},~\ref{quadmap} and~\ref{redshiftdep}). This implies that previous studies of the number of detectable haloes
have likely overestimated the number of foreground detections, at redshifts $z_{\rm h}<z_{\rm l}$. 
The exact magnitude of this overestimation appears to depend on the specific lensing 
configuration. As a reference, our experiments show this factor to be between 1.5 and 2 (see Fig.~\ref{nredshift}). 
These previous estimates are not based on a calculation of the Bayesian
evidence at $z_{\rm h}\neq z_{\rm l}$. Therefore, it is not currently
known whether an evidence-based criterion for detection would indeed
yield a dependence of the perturbers’ detectability on redshift that
is analogous to the one we measure. Certainly, we find that the
strategy adopted so far (of using deflection angles as proxies)
underestimates the degree of degeneracy in the lens modelling and
therefore artificially makes the detection of foreground perturbers 
easier {than in actually is in reality.} 

Concerning concentration, we find that detectability is a strong function of 
halo concentration, such that the population of detectable haloes is, in fact, a population
of systematically high-concentration haloes (see
Fig.~\ref{cshift}). The shift in the average concentration  
relative to the mass-concentration relation becomes increasingly
large for haloes of lower masses, and increases when a higher threshold
for detectability is adopted. For a threshold of $\DL_{\rm th}=50$,
the average shift in the concentration of haloes with masses below $10^{10}$~M$_{\rm \odot}$
is about $1.25\sigma_{\log c}$, where $\sigma_{\log c}$ is the lognormal scatter in the
mass-concentration relation. 

Crucially, accounting for the scatter in the 
mass-concentration relation results in a boost to the number of detectable haloes.
This boost is a strong function of the lensing configuration and of the threshold for detectability (or, equivalently, of the data quality as quantified by the maximum signal-to-noise; see Figs.~\ref{Nshift} and~\ref{Ncofn}).
As reference, for a combination of a lens configuration and  detection threshold 
that results in a total of 0.03 detections with $M_{\rm h} < 10^{10}$~M$_{\rm \odot}$ per lens in a CDM universe -- 
which is roughly comparable to what was previously predicted for lenses with HST data -- this boost 
amounts to a factor of $\approx 2.5$, and quickly grows to $\gtrsim10$ for the detections 
expected at $M_{\rm h} \lesssim 10^{9.5}$~M$_{\rm \odot}$.

Unfortunately, without a tailored study, it is impossible to provide a precise quantification 
of how the two effects above would combine to affect previous estimates of the expected number 
of detections in real observed strong lenses, especially since the two effects 
have opposite signs. The overestimate related to the redshift dependence is 
sensitive to the lensing configuration and, certainly to the redshifts of 
both lens and source, which here we have kept fixed. 
The underestimate due to the concentration dependence is 
a strong function of lensing configuration and data quality. It would appear that
the correction due to concentration is larger than that due to the redshift dependence, 
but further study is required to ascertain in which regime that is the case, and by 
how much.

What we can already establish in the present study is how
concentration effects can facilitate differentiating WDM models
with different cutoff halo masses. We have shown that taking into
account the dependence of the median halo concentration on the DM
model increases the spread among the number of expected detections
(see Fig.~\ref{Nofmhf}). For warmer models, lower halo concentrations
conspire with lower cosmological halo abundances increasingly to
suppress the number of detectable haloes. The effect of halo
concentration had not been included in previous studies, leaving only halo
abundances to differentiate among DM models, therefore
making it harder to distinguish them in strong lensing studies.

\begin{figure}
\centering
\includegraphics[width=\columnwidth]{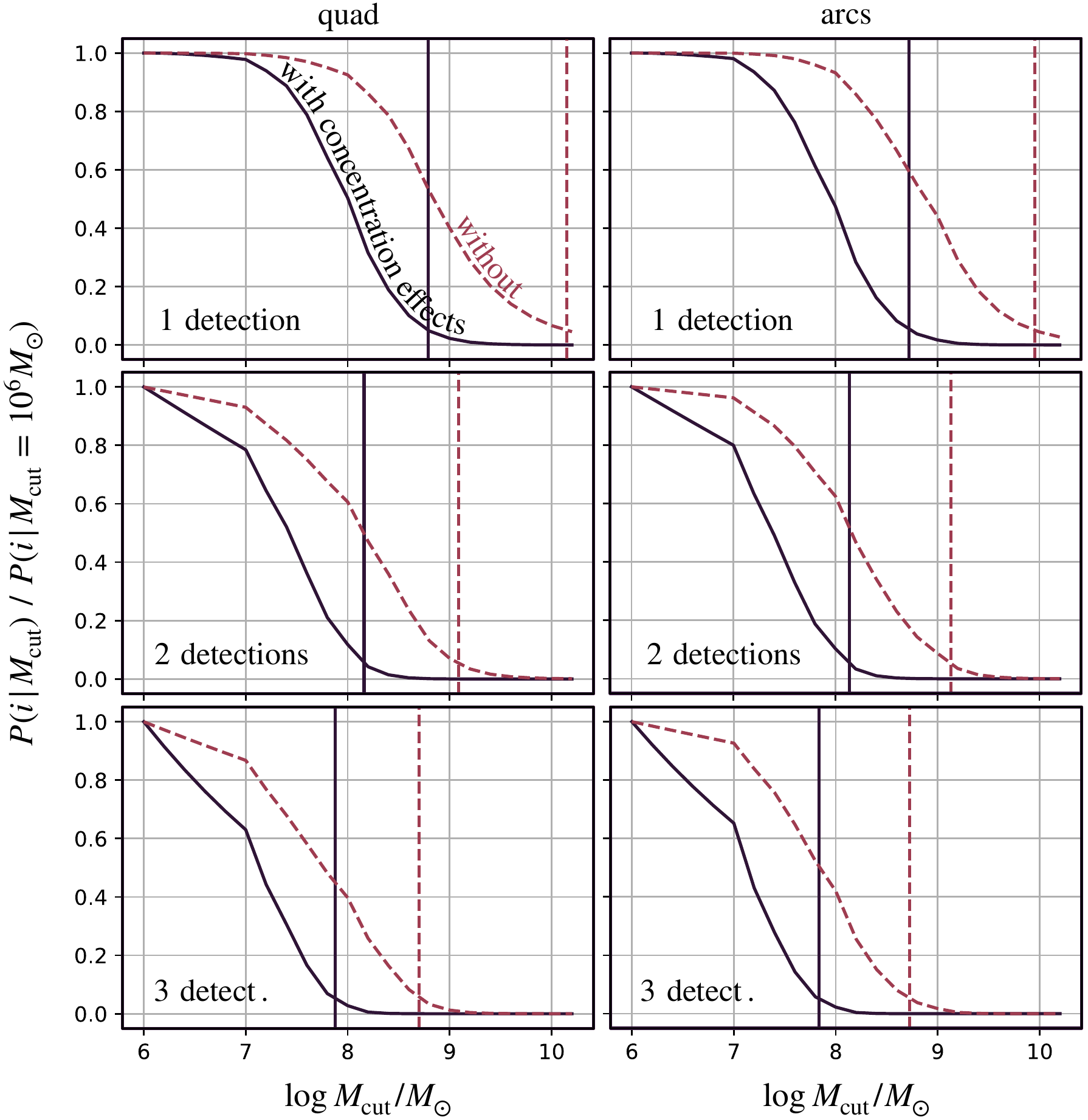}
\caption{The change on limits to the WDM cutoff mass, $M_{\rm cut}$,
  from including concentration effects, at a fixed number of expected detections for a CDM universe: $N_{\rm d, CDM}=1$.
Lines show likelihood ratios (see text) 
resulting from the detection of (1, 2, 3) perturbers, respectively,
from the top to the bottom
row. Dashed lines display the inference based on predictions that ignore concentration effects. These are included in the solid lines. }
\label{limits}
\end{figure}

\subsection{Sharper DM constraints from substructure lensing}
\label{final}

In order to quantify the extent to which concentration effects can sharpen 
future substructure lensing constraints, 
we assume we have a set of strong lenses such that the total expected number of detections in CDM is 
\begin{equation}
N_{\rm d, tot}(M_{\rm h}<10^{9.5}\,\textrm{M}_{\rm \odot}, {\rm CDM})
=1. 
\label{n1}
\end{equation}
We ignore the contribution of satellite haloes, and assume that the figure above
only includes haloes along the LOS, on which we have focussed in this work.
While we are not able to tailor our analysis quantitatively to any specific set of observed strong lenses, 
this figure is representative of what is achievable with current HST data \citep{Vegetti2018,Ritondale2019}, 
and therefore provides a useful reference point. 
We use our maps of log-likelihood increase, $\DL$, to calculate the number of expected detections in the 
same set of lenses for WDM models with different cutoff masses, $M_{\rm cut}$. 
We do so separately for our `quad' and `arcs' lensing configurations,
requiring that the number of lenses in the two separate sets  be such
as to satisfy Equation~\ref{n1} separately\footnote{
We require a detection threshold, $\DL_{\rm th}=30$, for our quad
configuration and $\DL_{\rm th}=20$ 
for our arcs configuration, which, as we have shown, leads to systematically fewer detections 
per lens.}. Furthermore, we set up predictions for both, the case in which concentration effects are accounted for
and the case in which they are ignored. Then, we compare the inferences on the DM model that would result
 from actually detecting $i = (1, 2, 3)$ individual haloes, in the two
 different configurations. These are displayed in Fig.~\ref{limits}, which
 shows the likelihood ratio, 
\begin{equation}
R = {{P( i | N_{\rm d, tot}(M_{\rm cut}))}\over {P( i | N_{\rm d, tot}(M_{\rm cut}=10^6 \, {\rm M_\odot}))}},
\label{likratio}
\end{equation}
where $P(\cdot|m)$ is the Poisson probability distribution with mean, $m$, and $i$ is the number of actual 
detections. Inference resulting from predictions that ignore
concentration effects are shown with dashed lines, while 
the likelihood ratio obtained when accounting for concentration
effects is shown with solid lines. The vertical 
lines indicate the limits on the WDM cutoff mass corresponding to a likelihood ratio of $R=0.05$. 
The right and left columns correspond, respectively, 
to the two lensing configurations.  Details are only marginally  different and the magnitude of the effect is very similar in the two cases:
at fixed number of expected haloes in a CDM universe, concentration effects make
constraints on the WDM cutoff mass significantly more stringent. The suppression in the 
concentration of WDM haloes enhances the effect of lower cosmological halo abundances,
allowing constraints that are about one order of magnitude more
stringent in $M_{\rm cut}$.

\subsection{Outlook}

Our results bring renewed confidence to the field of halo detection
with strong lensing data, and boost confidence that meaningful 
constraints can be obtained from analysis of current optical data. 
A number of previous works have contributed to the realization that it is extremely challenging
to use current optical HST data to obtain constraints on the
cutoff of the DM halo mass function that are competitive with those
obtained from the satellites of the Milky Way or  measurements of the
Lyman-$\alpha$ forest \citep[see][and references therein]{Enzi2021}. 
This is because, if halo abundances alone are used to differentiate
between CDM and WDM,
in order to be able to probe a WDM model with a cutoff mass, $M_{\rm
  cut}$ (the mass below which the abundance of haloes declines sharply), it is necessary to 
be sensitive to perturbers of that halo mass and below. 
However, evidence is mounting that detecting haloes of mass $M_{\rm h}\approx10^{8.5}$~M$_{\rm \odot}$ 
is extremely challenging with current lensing data, and therefore that
it would be very difficult to place competitive constraints. 

Concentration effects change this picture completely. For example, the limits displayed in Fig.~\ref{limits}
stem from detections of haloes of mass $M_{\rm h}>10^{9.5}$~M$_{\rm \odot}$ --
which is realistic with current HST data -- but they can rule out values
of the cutoff mass scale, 
$M_{\rm cut}\gtrsim10^8$~M$_{\rm \odot}$. This is a direct reflection of the effects of halo 
concentration, which, in contrast to halo abundances, first affects
haloes of masses significantly {\it above} the 
cutoff mass itself. For this reason, concentration effects allow substructure lensing studies to probe 
WDM models with cutoff masses at least one order of magnitude {\it below} the lowest sensitivity mass scale.
This implies that substructure lensing is, in fact, a much more
sensitive probe of the identity of  the DM 
than  had been previously recognized. 

\section*{Software Citations}

This work used the following software packages:
\begin{itemize}
\item 
\href{https://github.com/astropy/astropy}{\textt{Astropy}} 
\citep{astropy1, astropy2}
\item
\href{https://bitbucket.org/bdiemer/colossus/src/master/}{\textt{Colossus}}
\citep{colossus}
\item 
\href{https://github.com/matplotlib/matplotlib}{\textt{matplotlib}} 
\citep{matplotlib}
\item 
\href{https://github.com/numpy/numpy}{\textt{NumPy}} 
\citep{numpy}
\item 
\href{https://github.com/Jammy2211/PyAutoLens}{\textt{PyAutoLens}} 
\citep{Nightingale2015, Nightingale2018, Nightingale2021}
\item 
\href{https://www.python.org/}{\textt{Python}} 
\citep{python}
\item 
\href{https://github.com/scipy/scipy}{\textt{Scipy}}
\citep{scipy}
\end{itemize}

\section*{Acknowledgements}
NCA is supported by an STFC/UKRI Ernest Rutherford Fellowship, Project Reference: ST/S004998/1.
JWN and RJM acknowledge funding from the UKSA through awards ST/V001582/1 and ST/T002565/1; RJM is also supported by the Royal Society. ARR is supported by the European Research Council Horizon2020 grant `EWC' (award AMD-776247-6). CSF acknowledges support by the European Research Council (ERC) through Advanced Investigator grant to CSF, DMIDAS (GA 786910). RL acknowledge the support of National Nature Science Foundation of China (Nos 11773032,12022306).
This work used the DiRAC Data Centric system at Durham
University, operated by the Institute for Computational Cosmology
on behalf of the STFC DiRAC HPC Facility (www.dirac.ac.uk).
This equipment was funded by BIS National E-infrastructure capital 
grant ST/K00042X/1, STFC capital grants ST/H008519/1 and
ST/K00087X/1, STFC DiRAC Operations grant ST/K003267/1 and 
Durham University. DiRAC is part of the National E-Infrastructure.

\appendix

\section{Input macromodels}
Table~\ref{inputmacro} contains values of the adopted model parameters
for the two different lensing configurations used in this study.

\begin{table}
	\centering
	\caption{Input values for the two sets of macromodel parameters used in this study.}
	\label{inputmacro}
	\begin{tabular}{ccc} 
		\hline
		 Model parameter & quad & arcs \\
		\hline
		$(x_{\rm l},y_{\rm l})$ & (0.0051'', 0.0765'') & (0.0396'', 0.08'')  \\
		 $\epsilon_{\rm l}$& 1.165'' & 1.095''  \\
		 $\beta_{\rm l}$ & 1.93 & 1.99  \\
		 $(e_{1,\rm l},e_{2,\rm l})$ & (0.022, 0.011) & (-0.013, 0.007)  \\
		 $(\gamma_1,\gamma_2)$ & (-0.037, -0.099) & (-0.008, 0.001)  \\
		 $z_{\rm l}$ & 0.5 & 0.5 \\
		 $(x_{\rm s}, y_{\rm s})$ & (0.024'', 0.032'') & (-0.024, 0.036) \\
		 $r_{\rm eff}$ & 0.15'' & 0.12'' \\
		 $(e_{1,\rm s}, e_{2,\rm s})$ & (0.147, -0.135) & (0.05, -0.25) \\
		 $n_{\rm s}$& 1.1 & 1.2  \\
		 $I_{\rm s}$& 1 & 1  \\
		 $z_{\rm l}$ & 1.0 & 1.0 \\
		\hline
	\end{tabular}
\end{table}

\section{Dependence on the noise realization}

\begin{figure}
\centering
\includegraphics[width=.8\columnwidth]{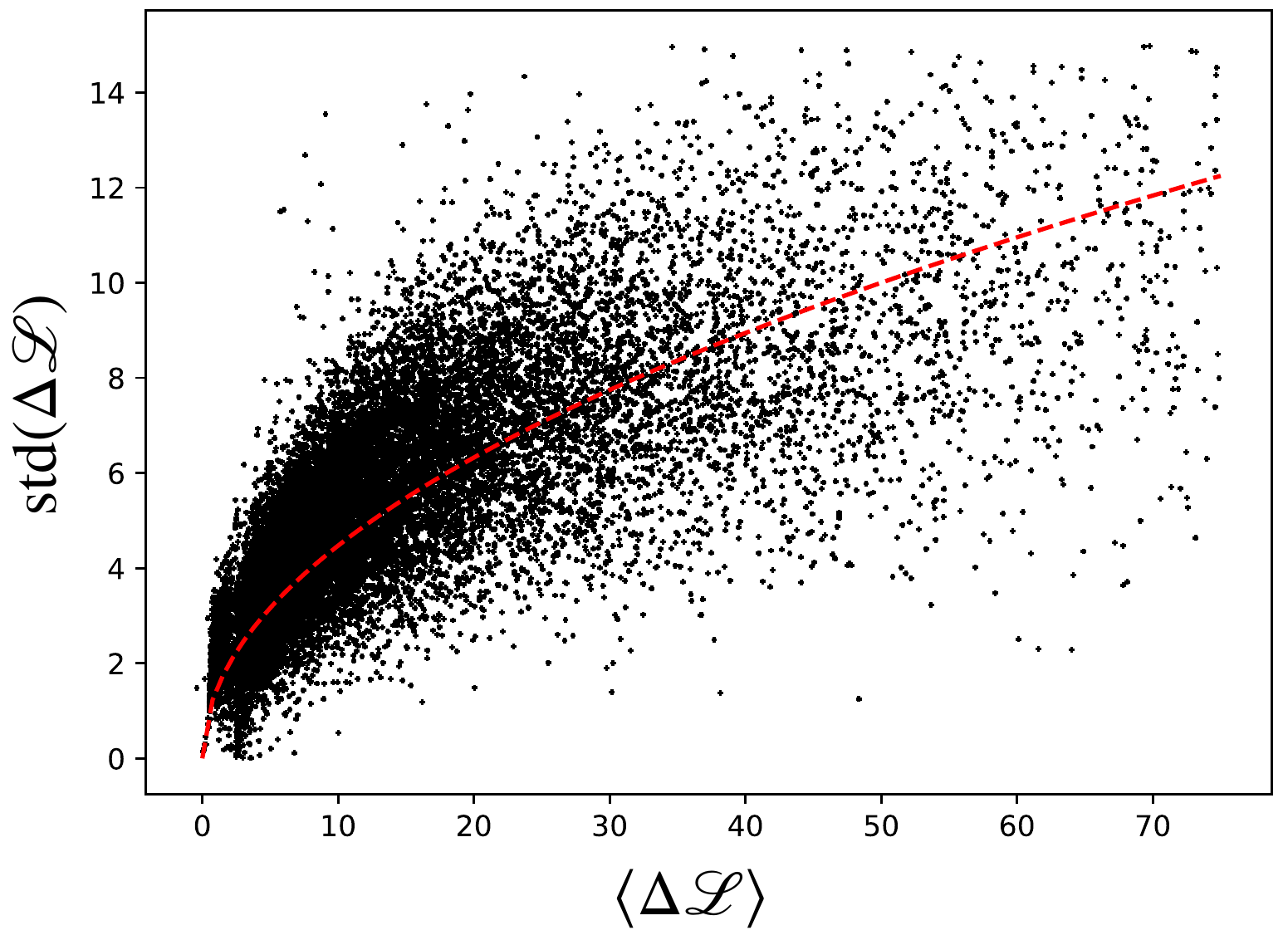}
\caption{The scaling of the scatter in the log-likelihood increase std$(\DL)$ resulting from different noise
realizations (at fixed signal-to-noise) and the mean log-likelihood increase $\langle \DL \rangle$. The 
red line shows the scaling of equation~(\ref{stdDL}). }
\label{mean:std}
\end{figure}

Fig.~\ref{mean:std} shows the link between the mean and the scatter (standard deviation)
of the log-likelihood gain $\DL$. Each point corresponds to a different set of halo properties
$\zh$. For each, 10 different random noise realizations have been considered and modelled. The 
corresponding set of values for the log-likelihood increase has been used to estimate
both mean value and standard deviation. The red dashed line illustrates the prediction 
of equation~(\ref{stdDL}).

\bsp	
\label{lastpage}

\end{document}